\documentclass[reprint,superscriptaddress,onecolumn,showpacs,amsmath,amssymb]{revtex4-1}
\usepackage{graphicx}
\usepackage{dcolumn}
\usepackage{bm}
\usepackage{epsfig}
\usepackage{color}
\usepackage{longtable}
\usepackage{float}
\usepackage{ulem}
\usepackage[french,english]{babel}
\usepackage[latin1]{inputenc}
\usepackage[OT1,T1]{fontenc}

\usepackage{longtable}
\usepackage{color}

\def\etal{{\it et al.}}                                        %

\begin{document}

\title{Theoretical Hyperfine Structures of $^{19}$F~I and $^{17}$O~I }

\author{Nouria Aourir}
\affiliation{Laboratoire d'\'Electronique Quantique, Facult\'e de Physique, USTHB, BP32,
El-Alia, Algiers, Algeria}

\author{Messaoud Nemouchi}
\affiliation{Laboratoire d'\'Electronique Quantique, Facult\'e de Physique, USTHB, BP32,
El-Alia, Algiers, Algeria}

\author{Michel Godefroid}
\affiliation{Chimie Quantique et Photophysique, Universit\'e libre de Bruxelles, B-1050 Brussels, Belgium}

\author{Per J\"onsson}
\affiliation{Materials Science and Applied Mathematics, Malm\"o University, S-20506 Malm\"o, Sweden}

\date{\today \\[0.3cm]}

\begin{abstract}
Multiconfiguration Hartree-Fock (MCHF)  and multiconfiguration Dirac-Hartree-Fock (MCDHF) calculations are performed for the $2p^{5}~^{2}P^{o}$,  $2p^4(^{3}P)3s~^{4}P$, $2p^4(^{3}P)3s~^{2}P$  and $2p^4(^{3}P)3p~^{4}S^o$ states of $^{19}$F~I  to determine their hyperfine constants. Several computing strategies are considered to investigate electron correlation and relativistic effects.  High-order correlation contributions are included in MCHF calculations based on single and double multireference  (SD-MR) expansions. The largest components of the single reference MCHF wave functions are  selected to define the MR sets. In this scheme, relativistic corrections are evaluated in the Breit-Pauli approximation. A similar strategy is used for the calculation of MCDHF relativistic wave functions and hyperfine parameters. 
While correlation and relativistic corrections are found to be rather small for the ground state, we highlight large relativistic effects on the hyperfine constant  $A_{3/2}$  of $2p^4(^{3}P)3p~^{4}S^o$ and, to a lesser extent, on $A_{1/2}$  of $2p^4(^{3}P)3s~^{4}P$. 
As expected for such a light system, electron correlation effects dominate over relativity in the calculation of the hyperfine interaction of all other levels considered. 
We also revisit the  hyperfine constants of  $2p^3(^{4}S)3s~^{5}S^{o}$  and  $2p^3(^{4}S)3p~^{5}P$ in 
$^{17}$O using  similar strategies. The results are found to be in excellent agreement with experiment.
 \end{abstract}


\maketitle

\section{Introduction} \label{sec:intro}

 $^{14,15}$N hyperfine structures have been investigated theoretically by J\"onsson \etal~\cite{Jonetal:2010a} revealing large discrepancies between theory and the results of the analysis of Doppler-free saturated absorption spectra by Jennerich \etal~\cite{Jenetal:06a}. Carette \etal~\cite{Caretal:10b} reconciled experiment with theory by a new interpretation of the observed signals and demonstrated that the apparition of crossover signals, if helpful in some cases \cite{Krietal:09a}, can be problematic and even completely misleading in the hyperfine structure assignments. The same Doppler-free saturated absorption spectroscopy technique was used for oxygen~\cite{JenTat:2000a} and fluorine \cite{TatAtu:97a}. Since these atoms are close neighbours of nitrogen in the periodic table, it is worthwhile to investigate if crossover signals did wrongly affect the analysis of their hyperfine spectra, by comparing experimental and theoretical hyperfine  constant values. Moreover, relativistic corrections can be larger than expected for theoretical hyperfine constants values of light elements~\cite{Caretal:13a} and it becomes crucial to estimate their magnitude by adequate computational strategies~\cite{CarGod:2011a}. \\

For $^{19}$F ($I=1/2$), the hyperfine coupling constant $A_{3/2}$ of the ground state $2p^5 \; ^2P^o_{3/2}$   has been determined from the analysis of the paramagnetic resonance absorption spectrum by Radford \etal~\cite{Radetal:61a}. Using the same technique, Harvey~\cite{Har:65a} extended this work by extracting the $A_{1/2}$ value of the metastable $^2P^o_{1/2}$ level of the inverted term.
At that time, the comparison of observation with theory was limited due to the poor quality of the best Hartree-Fock wave functions~\cite{Har:65a}. Things have evolved from the theoretical point of view with the pioneer development of computational methods taking electron correlation into account~\cite{GlaHib:78b}.
Doppler-free spectroscopy of excited states of atomic fluorine has been carried out for the very first time by Tate and Aturaliye~\cite{TatAtu:97a}. Hyperfine structure splittings and magnetic coupling constants of the upper and lower states of the $2p^4 (^3P) 3s \; ^2P_J - 2p^4 (^3P) 3p \; ^2D^o_{J'}$ transitions have been reported in this study, at least one order of magnitude more precise than values obtained using conventional spectroscopy \cite{Lid:49a,Hoc:78a} but in some cases, significantly different from the previously reported values. Levy \etal~\cite{Levetal:07a} performed hyperfine structure measurements via laser-induced fluorescence and modulated optical depopulation pumping for levels of the quartet spin system, i.e. for  $2p^4 (^3P) 3s \; ^4P_{5/2}$, $2p^4 (^3P) 3p \; ^4D^o_{5/2}$ and $2p^4 (^3P) 3p \; ^4D^o_{7/2}$. Very recently, Huo~\etal~\cite{Huoetal:2018a} determined the hyperfine constants of some levels within the  $2p^4 (^3P) 3s $, $2p^4 (^3P) 3p$ and  $2p^4 (^3P) 3d$ configurations, using high resolution absorption spectroscopy. 
{\it Ab initio} calculations of hyperfine structures remain scarce for neutral light systems for which relativistic corrections on the electronic atomic structures are {\it a priori} expected to be smaller than electron correlation effects. For fluorine however ($Z=9$), the relativistic corrections to the nonrelativistic hyperfine parameters have been shown to be surprisingly large (around 30\%) for the $A$ values of 
$2p^4 (^3P) 3p \; ^4P^o_{3/2}$ and $2p^4 (^3P) 3p \; ^4P^o_{5/2}$~\cite{Caretal:13a}. 
These effects, predicted theoretically, were confirmed  by the recent measurements of Huo~\etal~\cite{Huoetal:2018a}. For example, the nonrelativistic value of $A(2p^4 (^3P) 3p \; ^4P^o_{3/2}$)$=1374$~MHz that was found to increase to 1784~MHz due to relativity, approaches the experimental value 1824(20)~\cite{Huoetal:2018a}.
As an extension of the theoretical study by Carette \etal~\cite{Caretal:13a} who investigated relativistic effects on the hyperfine structures of the odd terms
$2p^4 (^3P) 3p \; ^2D^o$, $2p^4 (^3P) 3p \; ^4D^o$ and $2p^4  (^3P) 3p \; ^4P^o$, we consider in the present work the ground term $2p^5 \; ^2P^o$ and the excited terms, $2p^4 (^3P) 3s \; ^4P$,  $2p^4  (^3P) 3s \; ^2P$    and  $2p^4(^{3}P)3p~^{4}S^o$  for which experimental magnetic coupling constants have been determined, unfortunately not for all $J$-levels~\cite{Radetal:61a,Levetal:07a,TatAtu:97a}. \\



For $^{17}$O ($I=5/2$), Marin \etal~\cite{Maretal:93a} investigated the hyperfine structures in the $3s - 3p$ transitions of the triplet and quintet spin systems, using high-resolution saturation spectroscopy. 
The hyperfine structure coupling constants of the four levels involved in the $2p^3 (^4D) 3s \; ^5S^o_{2}$ $-$ $2p^3 (^4D) 3p \; ^5P_{1,2,3}$ transitions have been determined by  Jennerich and Tate~\cite{JenTat:2000a}, with the same technique, but claiming a higher precision than \cite{Maretal:93a}. Pioneer nonrelativistic multiconfiguration calculations were performed by Godefroid \etal~\cite{Godetal:97b} for $3s \; ^5S^o_{2}$ and by J\"onsson and Godefroid \cite{JonGod:00a} for these low-lying levels. Elaborate multiconfiguration methods allowing the simultaneous inclusion of electron correlation and relativity have been developed during the last two decades~\cite{Froetal:2016a}. Their availability give new impetus
for revisiting the hyperfine structure of these levels  to clarify remaining theory-observation discrepancies  in the same line than fluorine. \\

The multiconfiguration approaches that we use in both nonrelativistic and relativistic schemes are fully described in \cite{Froetal:2016a}. The underlying variational formulation of these two theories differ in detail, but is very similar in concept. We will therefore limit Section~\ref{sec:theory} to the basic equations needed to understand the physical content of the hyperfine parameters.

\clearpage

\section{Theory}
\label{sec:theory}

\subsection{Hyperfine Interaction}

The hyperfine interaction in atomic systems is due to the interaction between the electrons and the electromagnetic multipole moments of the nucleus. The two lowest multipole orders of the hyperfine interaction are represented in the Hamiltonian by the two following contributions
\begin{equation}
  H_{\rm{hfs}}= \bm {T}^{(1)} \cdot \bm{M}^{(1)} + \bm {T}^{(2)} \cdot \bm{M}^{(2)},
\end{equation}
that describe, respectively, 
the dipole magnetic interaction and the quadrupole electric interaction. 
In the purely relativistic sheme the magnetic  electronic tensor operator for an $N$-electron system is (in atomic units) given by~\cite{LINROS:75a}
\begin{equation}
\bm {T}^{(1)} = - \sqrt{-1} \; \alpha \sum_{i=1}^{N}\left(\bm \alpha_i \cdot \; \bm l_{i}\bm C^{(1)}(i)\right)r_{i}^{-2},
\end{equation}
whereas in the nonrelativistic limit, it is written as a sum of three contributions

\begin{equation}
\bm {T}^{(1)}=\bm {T}^{(1)}_{l}+\bm {T}^{(1)}_{sd}+\bm {T}^{(1)}_{c} \; .
\end{equation}
$\bm {T}^{(1)}_{l}$,  $\bm {T}^{(1)}_{sd}$ and $\bm {T}^{(1)}_{c}$ are the magnetic fields produced at the site of the nucleus, due to, respectively, the orbital motion of the electrons (orbital term), the spin precession of electrons (spin-dipolar term) and the penetration of electrons inside the nuclear volume (contact term).
The electric quadrupole tensor operator 
keeps the same form in both nonrelativistic and relativistic schemes and writes as (in atomic units)
\begin{equation}
\bm {T}^{(2)} = -\sum_{i=1}^{N} \bm C^{(2)}(i) r_{i}^{-3} \; .
\end{equation}
The three different contributions to the nonrelativistic hyperfine constant $A_J$ are written as~\cite{Hib:75b,Jonetal:93a}
\begin{equation}
A_{J}^{orb}=G_{\mu}\frac{\mu_{I}}{I}\;a_{l}\frac{\langle\,\vec{L}.\vec{J}\rangle}{LJ(J+1)} \; ,
\label{Aorb}
\end{equation}

\begin{equation}
A^{sd}_{J}=\frac{1}{2}\,G_{\mu}\,g_{s}\,\frac{\mu_{I}}{I}\,a_{sd}\frac{3\,\langle\,\vec{L}.\vec{S}
\,\rangle\,\langle\,\vec{L}.\vec{J}\,\rangle\,-\,L(L+1)\,\langle\,\vec{S}.\vec{J}\,\rangle}{SL(2L-1)J(J+1)}
\; ,
\label{Asd}
\end{equation}

\begin{equation}
A_{J}^{c} = \frac{1}{6}\,G_{\mu}\,g_{s}\,\frac{\mu_{I}}{I}a_{c}\,
\frac{\langle\,\vec{S}.\vec{J}\rangle\,}{SJ(J+1)} \; ,
\label{Ac} 
\end{equation}

\begin{equation}
B_{J} =-G_{q}\,Q\,b_{q}\,\frac{6\langle\,\vec{L}.\vec{J}\,\rangle^{2}\,-\,3\langle\,\vec{L}.\vec{J}\,\rangle\,-\,2L(L+1)J(J+1)}{L(2L-1)(J+1)(2J+3)} \; ,
\end{equation}
where $g_s = 2.0023193$ is the electronic $g$ factor corrected for the quantum electrodynamic (QED) effects. The $J$-dependent parameters  $a_{l}$, $a_{sd}$, $a_{c}$, $b_{q}$ are proportional to reduced matrix elements of specific operators

\begin{equation}
a_{l}=\sqrt{\frac{L}{(L+1)(2L+1)}}\;\; \langle \gamma LS
\|\sum_{i=1}^{N}\bm {l}^{(1)}_{i}r_{i}^{-3}\|\gamma LS \rangle \; ,
\label{al}
\end{equation}

\begin{equation}
a_{sd}=\sqrt{\frac{LS(2L-1)}{(L+1)(2L+1)(2L+3)(S+1)(2S+1)}} \langle \gamma LS \|\sum_{i=1}^{N}\bm
{C}^{(2)}_{i}s^{(1)}_{i}r_{i}^{-3}\|\gamma LS \rangle \; ,
\label{asd}
\end{equation}

\begin{equation}
a_{c}=\sqrt{\frac{S}{(S+1)(2S+1)}}\;\; \langle \gamma LS
\|\sum_{i=1}^{N}2\bm {s}^{(1)}_{i}\delta(r_{i})r_{i}^{-2}\|\gamma 
LS \rangle \; ,
\label{ac}
\end{equation}

\begin{equation}
b_{q}=\sqrt{\frac{L(2L-1)}{(L+1)(2L+1)(2L+3)}}\,\langle \gamma LS
\|\sum_{i=1}^{N}2\bm {C}^{(2)}_{i}r_{i}^{-3}\|\gamma LS \rangle \; .
\label{bq}
\end{equation}
$G_{\mu}= 95.41068$  and $G_q = 234.9647$ are the numerical factors needed to get the $A$ and $B$ constants in MHz when expressing $\mu_I$ in nuclear magnetons, $Q$ in barns ($10^{-28}$~m$^2$) and the hyperfine parameters in atomic units ($a_0^{-3}$). 

\subsection{Many-electron wave functions}

\subsubsection{Multiconfiguration Hartree-Fock approach}
In the multiconfiguration Hartree-Fock (MCHF) method the nonrelativistic wave function  $\Psi$, for a state labelled $\vert \alpha LSM_L M_S \pi \rangle$, is expanded in terms  of $N_c$ configuration state functions (CSF) $\Phi$, which are eigenfunctions of $\bf L^2$, $\bf S^2$, $L_z$,  $S_z$ and the inversion (parity) operator $I$, namely: 
\begin{equation}
\label{MCHF_wfn}
\Psi (\alpha  L S M_L M_S  \pi ) = \sum_{k=1}^{N_c} c_k \; \Phi_k( \alpha_k L S M_L M_S \pi)  \; .
\end{equation}
The CSFs are built on a basis of one-electron spin-orbital functions
\begin{equation}
\label{nr_spin_orbitals}
\phi_{nlm_lm_s}({\bf r}, \sigma ) = \frac{P_{nl}(r)}{r} Y _{l m_l} (\theta, \varphi) \chi_{m_s} (\sigma) \; .
\end{equation}
 In the MCHF procedure
{\it both} sets of
radial functions $\{ P_{n_i l_i} (r) \} $ and mixing coefficients $\{ c_k \}$,
are optimized to self-consistency by solving
numerically and iteratively the multiconfiguration Hartree-Fock differential
equations
for the former and the configuration interaction (CI) problem for the latter \cite{FBJ:97a,Froetal:2007a}. 
In a CI calculation, the mixing coefficients are determined by 
diagonalising the Hamiltonian matrix in a CSF basis built on a set of pre-optimised one-electron orbitals. This approach allows to investigate in a systematic way the extension of the configuration space using MCHF orbitals~\cite{Froetal:2016a}.

\subsubsection{Multiconfiguration Dirac-Hartree-Fock approach}
\label{ssec:MCDHF}
In the Multiconfigurational Dirac-Hartree-Fock (MCDHF) method,  the atomic state is described by a wave function, eigenfunction of    $\bf J^2$, $J_z$, parity $\pi$, written as a relativistic CSF expansion
\begin{equation}
\label{MCDHF_wfn}
\Psi (\alpha  JM_J\pi) = \sum_{k=1}^{N_c} c_k \; \Phi_k( \alpha_k JM_J\pi). 
\end{equation}
The CSFs in this case are symmetry-adapted linear combinations of Slater determinants built on one-electron Dirac orbitals
\begin{equation}
\label{Dirac_spinor}
\phi_{n \kappa m} ({\bf r}, \sigma )  = \frac{1}{r} \left( 
\begin{array}{c}
P_{n \kappa} (r) \; \chi_{\kappa m} (\theta , \varphi) \\
i Q_{n \kappa} (r) \; \chi_{- \kappa m} (\theta , \varphi) 
\end{array}
\right) \; .
\end{equation}
where $i$ multiplying the small component radial function $Q_{n \kappa}$ is the imaginary unit defined by $i^2 =-1$.
The radial functions $\{P_{n_i \kappa_i} (r)\}$ and $ \{Q_{n_i \kappa_i} (r) \}$ and the mixing coefficients 
$\{c_k\}$  are obtained by solving iteratively, until self-consistency,  the multiconfiguration Dirac-Hartree-Fock differential equations for the one-electron radial functions and the CI diagonalization problem for the mixing coefficients. The Dirac-Coulomb Hamiltonian $H_{DC}$ used to derive the MCDHF equations writes as
\begin{equation}
\label{eq:DC_Ham}
H_{\rm{DC}}=\sum_{i=1}^{N}~h_{\rm{D}}(i)+\sum_{j>i=1}^{N}\frac{1}{r_{ij}} \; ,
\end{equation}
where  $h_{\rm{D}}$ is the one-electron Dirac operator
\begin{eqnarray}
\label{Di_Ham}
h_{D}(i)=  c\, {\bm{ \alpha }}_i \cdot {\bm{ p }}_i + (\beta_i -1)c^2 + V^{nuc}_i  \; .
\end{eqnarray}

As in the nonrelativistic approach, a configuration interaction calculation can be performed in a given CSF basis built on pre-optimised one-electron Dirac spinors. The corresponding relativistic calculations are labeled RCI in the following. For these RCI calculation, one can use 
the many-electron  Dirac-Coulomb-Breit Hamiltonian in the long wavelength approximation
\begin{equation}
H_{\rm{DCB}} \simeq H_{\rm{DC}} + H_{\rm{Breit}}  \; ,
\end{equation}
where $H_{\rm{Breit}}$ is the Breit interaction
\begin{equation}
H_{\rm{Breit}}= -\sum_{j>i=1}\frac{1}{2r_{ij}}{\left[\bm{\alpha}_{i} \cdot
  \bm{\alpha}_{j}+\frac{(\bm{\alpha}_{i} \cdot \bf {r}_{ij})
  (\bm{\alpha}_{j}\cdot\bf {r}_{ij})}{r_{ij}^{2}}\right]} \; ,
\end{equation}
 to which further QED corrections such as self-energy and vacuum polarization can be added~\cite{Froetal:2016a}.  \\

\subsubsection{Relativistic calculations with MCHF orbitals}

\paragraph{Breit-Pauli approximation}

In the Breit-Pauli (BP) approximation, relativistic effects are accounted for by adding to the nonrelativistic Hamiltonian additional terms of order $\alpha^2$ to approximate $H_{\rm{DCB}}$ \cite{ArmFen:74a}. These relativistic operators are included in the BP Hamiltonian matrix built in a $(LS)J$ basis of CSFs where ${\bf J}$ results from the angular momenta coupling ${\bf J} = {\bf L} + {\bf S}$. The atomic state function  is then described by a 
Breit-Pauli eigenvector
\begin{equation}
\label{BP_wfn}
\Psi (\alpha  J M_J \pi ) = \sum_{k=1}^{N_c} c_k \; \Phi_k( \alpha_k L_k S_k J M_J \pi ) \; 
\end{equation} 
allowing $LS$ mixing due to the fine-structure BP terms, i.e. the spin-orbit, the spin-other orbit and the spin-spin corrections, that do not commute with ${\bf L}$ and ${\bf S}$.
In this Breit-Pauli CI approach, the CSFs $\Phi_k( \alpha_k L_k S_k J M_J \pi )$ are built on nonrelativistic radial one-electron functions pre-optimised through MCHF calculations. \\ 

\paragraph{The RCI-P approach}
In the nonrelativistic limit ($c \rightarrow \infty$) known as the Pauli approximation, the small component of the one-electron Dirac orbital can be estimated from the large one, as 
\begin{equation}
Q_{n\kappa}(r)  \simeq  \frac{\alpha}{2} \left( \frac{d}{dr} + \frac{\kappa}{r} \right) P_{n \kappa}(r) 
\;.
\end{equation}
A RCI-P calculation consists in using the RCI relativistic configuration interaction approach described above (see section~\ref{ssec:MCDHF}), with wave function expansions of the kind (\ref{MCDHF_wfn}) built on Dirac spinors (\ref{Dirac_spinor}) and using the Pauli approximation to estimate the small component radial functions from  nonrelativistic MCHF radial orbitals:
\begin{eqnarray}
\label{Pauli_approx}
P_{n\kappa}(r) &=& P_{nl}^{\rm{MCHF}}(r) \; ,\nonumber \\
Q_{n\kappa}(r) & \simeq & \frac{\alpha}{2} \left( \frac{d}{dr} + \frac{\kappa}{r} \right) P^{\rm{MCHF}}_{n l}(r) \; .
\end{eqnarray}

\section{Building the CSF spaces}

The nonrelativistic MCHF, CI, and MCHF+BP calculations are performed with the ATSP2k package~\cite{Froetal:2007a} while the relativistic MCDHF, RCI(-P) calculations are realised with GRASP2K~\cite{Jonetal:2013a}. The MCHFMCDHF program converts the MCHF orbitals into Dirac spinors using the Pauli approximation for the RCI+P calculations. All these calculations require the definition of specific building rules to generate the CSF expansions (\ref{MCHF_wfn}), (\ref{MCDHF_wfn}) or (\ref{BP_wfn}) from a given orbital active set (AS). These rules are not arbitrary and respect the desired symmetries.
For all calculations, the wave function expansions  are obtained with the active space method, where the CSFs are generated by one or more excitations of electrons to an active set of orbitals, from a single reference (SR) or from a set of CSFs defining a multireference (MR) space. The latter chosen to capture the dominant correlation excitations in a zeroth-order picture~\cite{Froetal:2016a}. \\

Several methods of constructing configuration spaces are considered in the present work. For a given method, the orbital active space (AS) is characterised by $[n_{max}]$ when no angular limitation applies, and by $[n_{max} l_{max}]$ if some angular orbital limitation is introduced.
 In order to highlight the electron correlation effects, the hyperfine structure constants are monitored as functions of the AS.


\subsection{ Single-reference method} 
In a single-reference (SR) calculation, the configuration space is generated by allowing single- (S) and double- (D) excitations from the reference CSF.  In the present work, the SR configurations are 
$1s^2 2s^2 2p^5$, $1s^2 2s^2 2p^4 3s$  and $1s^2 2s^2 2p^4 3p$ for fluorine, and $1s^2 2s^2 2p^3 3s$ and  $1s^2 2s^2 2p^3 3p$
for oxygen. 
In the following, the MCHF, BP, RCI and MCDHF calculations performed using this single-reference approach are respectively labelled SR-MCHF, SR-BP, SR-RCI and  SR-MCDHF. For BP calculations, all allowed $LS$-symmetries for a given $J$-value are included in the CSF-expansion~(\ref{BP_wfn}). 

\subsection{ Multi-reference method} 
In the multi-reference (MR) approach, a zeroth-order set of configurations is selected on the basis of the largest cumulative weight, $w_k=\sqrt{\sum_i (c_i^k )^2}$, calculated from the multiconfiguration SR-MCHF[10g] eigenvectors. The summation $\sum_i$ appearing in the definition of $w_k$ runs over all CSFs $i$ that belong to configuration~$k$, with all possible coupling trees. The six most important configurations obtained by sorting the corresponding eigenvector's components according to their weights $w_k$, are reported in Table~\ref{tab:weights} for all states considered in F~I and O~I. Once this sorting done, different MR$x$ subsets containing the $x$ largest configuration component can be defined, with $ 2 \le x \le 6$.
In the SD-MR computational strategy, SD excitations from  each member of a  MR subset are included, with the restriction that excited CSFs $\Phi_i$ will be kept in the SD-MR$x$ expansion if and only if they interact with at least one of the CSF of the MR$x$ set, i.e. 
\begin{equation} \Phi_i \in \text{SD-MR$x$-expansion} \;\;\; \Leftrightarrow ~\exists \;\; \{\Phi_k\} \in \text{MR}
x ,  \;\;\; \text{with} \;\; \langle \Phi_k |H|\Phi_i\rangle \ne 0 \; , 
\end{equation} 
where $H$ is the nonrelativistic Hamiltonian of Schr\"odinger or the Dirac-Coulomb Hamiltonian $H_{\rm{DC}}$ of (\ref{eq:DC_Ham}).
\begin{table}[H]
\caption{List of the six configurations having the largest weight $w_k$ (see text) arising from SR-MCHF[10g] calculations on  $2p^3(^4S)3s \; ^5S^o$ and  $2p^3(^4S)3p \; ^5P$ of O~I \ and    $2p^5 \; ^2P$, $2p^4(^3P)3s \; ^4P $, $2p^4(^3P)3s \; ^2P $  and  $2p^4(^3P)3p \; ^4S $ of F~I.}
\begin{center} 
\begin{tabular}{llcccllc} 
 \\
 \hline \hline \\
 [-0.2cm]
                                             \multicolumn{3}{c}{Oxygen}                                &&& \multicolumn{3}{c}{Fluorine}\\
\hline
Term                                     &  Configuration                       &     $w_k$            &&&      Term                               &  Configuration                   &     $w_k$                         \\
\hline
                                         
                                            &                                                 &                       &&&                                     &                                               &                              \\
                                              & 1.~$2s^2 2p^3 3s  $                  &  0.9847          &&&                                     & 1.~$ 2s^2 2p^5$                         &  0.9810                \\
                                            & 2.~$2s 2p^3 3s  3d $              &  0.1093          &&&                                     & 2.~$2s^2 2p^3 3p^2$                 &  0.1026        \\                                               
$2p^3(^4S)3s \; ^5S^o$      & 3.~$2s  2p^2 3s  3p  4s $  &  0.0502          &&&  $2p^5 \; ^2P^o$              & 3.~$2s^2 2p^3 3d^2$                 &  0.0809         \\
                                            & 4.~$2s^2 2p  3s  3p^2$          &  0.0450          &&&                                      & 4.~$2s  2p^4 3s  3p $        &  0.0693         \\
                                            & 5.~$2s^2 2p  3s  3d^2$          &  0.0423          &&&                                      & 5.~$2s  2p^5 3d $                 &  0.0638         \\
                                            & 6.~$2s  2p^4 3p $                   &  0.0414          &&&                                     & 6.~$2s^2 2p^4 3p $                 &  0.0361          \\
                                            &                                                 &                       &&&                                     &                                               &                       \\
                                            & 1.~$2s^2 2p^3 3p $                   &  0.9857          &&&                                     & 1.~$2s^2 2p^4 3s $                  &  0.9844          \\
                                            & 2.~$2s  2p^3 3p  3d $          &  0.1102          &&&                                      & 2.~$2s  2p^4 3s  3d $         &  0.1939          \\  
                                            & 3.~$2s  2p^2 3s  3p  4p $  &  0.0537         &&&                                      & 3.~$2s^2 2p^2 3s  3p^2$          &   0.0685         \\                                         
$2p^3(^4S)3p \; ^5P$          & 4.~$2s  2p^3 3s  3p $          &  0.0467          &&&  $2p^4(^3P)3s \; ^4P$ &  4.~$2s^2 2p^2 3s  3d^2$          &  0.0612         \\ 
                                            & 5.~$2s^2 2p  3p  3d^2$          &  0.0425          &&&                                      & 5.~$2s  2p^3 3s  3p  4s $  &  0.0530        \\
                                            & 6.~$2s^2 2p  3p  4p^2$          &  0.0416          &&&                                      & 6.~$2s^2 2p^3 3s  3p $           &   0.0364                     \\
                                            &                                                &                        &&&                                      &                                                 &                        \\
                                            &                                                &                        &&&                                      & 1.~$2s^2 2p^4 3s $                    &  0.9847              \\
                                            &                                                &                        &&&                                      & 2.~$2s  2p^4 3s  3d $           &   0.0860                     \\       
                                            &                                                &                        &&&  $2p^4(^3P)3s \; ^2P $ & 3.~$2s^2 2p^2 3s  3p^2$           &       0.0683                 \\         
                                            &                                                &                        &&&                                      & 4.~$2s^2 2p^2 3s  3d^2$           &     0.0612               \\
                                            &                                                &                        &&&                                      & 5.~$2s  2p^3 3s  3p  4s $   &     0.0451                     \\   
                                            &                                                &                        &&&                                      & 6.~$2s^2 2p^3 3s  3p $            &      0.0399                         \\  
                                            &                                                &                        &&&                                      &                                                       &                        \\
                                            &                                                &                        &&&                                      & 1.~$2s^2 2p^4 3p $                    &   0.9856              \\
                                            &                                                &                        &&&                                      & 2.~$2s  2p^4 3p  3d $           &    0.0863                     \\       
                                            &                                                &                        &&&  $2p^4(^3P)3p \; ^4S^o $ & 3.~$2s^2 2p^2 3p  4p^2$           &    0.0650                 \\         
                                            &                                                &                        &&&                                      & 4.~$2s^2 2p^2 3p  3d^2$           &    0.0615               \\
                                            &                                                &                        &&&                                      & 5.~$2s  2p^3 3s  3p  4p $   &    0.0556                     \\   
                                            &                                                &                        &&&                                      & 6.~$2s^2 2p^3 3p  4p $            &    0.0304                         \\ 
 &                                                 &                       &&&                                     &                                               &                              \\                                                                                       
 \hline \hline \\
\end{tabular}
\end{center}
\label{tab:weights}
\end{table}

\section{Results and discussion}
\subsection{Fluorine}
\subsubsection{$2p^5 \; ^2P^o_{1/2,3/2}$ and $2p^4(^3P)3s \; ^2P_{1/2,3/2} $}

The hyperfine constants $A_J$ are reported in Table~\ref{tab:A_J2P2P} for the  ground term levels $2p^5 \; ^2P^o_{1/2,3/2}$  and for the excited states $2p^4(^3P)3s \; ^2P_{1/2,3/2} $. These constants have been estimated using $I=1/2$ and the magnetic dipole nuclear moment value $\mu (^{19}\mbox{F}) = 2.628868$ nuclear magneton (nm) taken from Stone's compilation~\cite{Sto:2005a}.
Starting from the single-configuration Hartree-Fock (HF) approximation, we performed single and double SR-MCHF calculations by extending systematically the orbital AS up to [10g]. The multireference MCHF calculations  (MR4-MCHF) correspond to SD excitations from the four first configurations appearing in the lists of table~\ref{tab:weights} to the same [10g] AS. These expansions become too large for  relativistic calculations. The MR4-BP  expansion were therefore limited to SD excitations to [7g] for the third and fourth configurations of the list but keeping the [10g] AS for the first two.  This is noted in Table~\ref{tab:A_J2P2P}  as MR4-BP[10g,10g,7g,7g], where [AS1,AS2,AS3,AS4] represent the four active sets used respectively for the four reference configurations included in MR4. Using the same notation, the MR4-MCDHF and MR4-RCI expansions use, respectively the [10g,5f,5f,5f] and [10g,5g,5g,5g] active sets, where the MR4-RCI calculation is a relativistic CI calculation using the MR4-MCDHF radial orbitals. 

\begin{table}[H]
\caption{Hyperfine structure constants $A_J$ (in MHz) of $2p^5 \; ^2P^{o}_{1/2,3/2}$ and $2p^4(^3P)3s \; ^2P_{1/2,3/2} $ of $^{19}$F~I. $N_c$ is the number of CSFs. AS specifies the orbital active set (see text for the notations).}
\begin{center}
\begin{tabular}{llccccccc}\\ 
\hline \hline \\[-0.2cm]
                   &                       & \multicolumn{3}{c}{$2p^5~^2P^o$} &    \multicolumn{4}{c}{$2p^4(^3P)3s \; ^2P $}  \\
Method       &  AS                 & $N_c$ &       $A_{1/2}$    &  $A_{3/2}$       & & $N_c$&      $A_{1/2}$      &    $A_{3/2}$              \\ [0.1cm]
\hline  \\[-0.2cm]
                    \multicolumn{9}{c}{nonrelativistic} \\
 [0.2cm]                   
HF              &                        &     1       &10099.08  & 2018.06&& 1& 2077.27 & 3235.20 \\                   
SR-MCHF  &      [10g]          & 12912 & 10271.61 &1974.79  & & 36581&1711.21   &3096.34 \\       
                   &                        &            &                 &               & &           &               &                       \\
MR4-MCHF  & [10g]            &244639& 10194.28 & 2012.07  & &274316 & 1655.93&3098.89  \\
                   \hline  \\[-0.2cm]  
            \multicolumn{9}{c}{relativistic} \\ 
[0.2cm]            
   SR-BP     & [10g]    & 74902 &10361.53  &1975.99    & &271413&  1834.22   &3030.35\\ 
SR-RCI-P        &   [10g]   &111720 &10344.56  &1974.13    &&291032  &1833.41     & 3028.26\\ 
SR-MCDHF   & [10g]   & 111720  &10350.43  &1971.54 &&291032   & 1840.42  &3026.27\\     
                   &            &            &                  &               &&     &                &              \\
MR4-BP     &  [10g,10g,7g,7g]   & 826947&10283.74  &2012.38   &&689301   & 1774.85  &3036.90 \\ 
MR4-MCDHF & [10g,5f,5f,5f]   & 350905  &10282.81  &2004.67 &  & 450703  & 1792.39  &3038.99\\
MR4-RCI  & [10g,5g,5g,5g]   & 615148  &10278.47  &2005.48 & & 456781  & 1775.16  &3036.62\\
\hline \\[-0.2cm]
                   \multicolumn{9}{c}{Other theory} \\
[0.2cm]                   
Glass and Hibbert~\cite{GlaHib:78b} &       && 10210.9  &  2014.1 & &&& \\
\hline  \\[-0.2cm]
 \multicolumn{9}{c}{Observed} \\
 [0.2cm]
\multicolumn{3}{l}{Radford \etal~\cite{Radetal:61a}}   &     & $ 2009.99(1) $& &&&\\
\multicolumn{3}{l}{Harvey~\cite{Har:65a}}       & $ 10244.21(3) $& &&&& \\
\multicolumn{3}{l}{Tate and Aturaliye~\cite{TatAtu:97a}}&&    &   &  &$1737.1(4)$ & $3057.9(21) $\\
\multicolumn{3}{l}{Huo \etal~\cite{Huoetal:2018a}}      &&    &   &  &$1550(200)$ & $3048(10) $\\
\\[0.1cm]
\hline \hline \\
\end{tabular}
\end{center}
\label{tab:A_J2P2P}
\end{table}
As shown in Table~\ref{tab:A_J2P2P}, the  MR4-BP values are in satisfactory agreement with observation~\cite{Radetal:61a,Har:65a} and with the pioneer theoretical predictions of Glass and Hibbert~\cite{GlaHib:78b} for the ground configuration. Some reassuring observations can be done: the SR-RCI-P and SR-BP values agree well with each other. The  SR-MCDHF and  MR4-RCI results are consistent with respectively, the SR-BP and  MR4-BP values. For the excited $2p^4(^3P)3s \; ^2P$ term, the hyperfine constant values calculated with the three relativistic approaches (MR4-BP, MR4-MCDHF and MR4-RCI) are consistent and agree reasonably well  with the experimental value of Tate and Aturaliye~\cite{TatAtu:97a} and also with the most recent, but less accurate, values of Huo \etal~\cite{Huoetal:2018a}. 
Let $\Delta(a,b) = (a-b)/a$ be the relative difference between non-zero values $a$ and $b$. If $b$ is our theoretical result and $a$ is  the experimental value we can consider $\Delta(a,b) $ as relative uncertainty  estimates of our results. We then estimate the relative uncertainty on the ground state hyperfine constants around 0.3\% and 2\% and 0.7\%, respectively for  $A_{1/2}$ and $A_{3/2}$ of the excited state.\\

We report in Table~\ref{ER}  those $\Delta(a,b)$ values (in \%) for the hyperfine constants estimated with different theoretical models.
   The analysis of these relative differences for the $A_{1/2}$ and $A_{3/2}$ of the ground term levels $2p^5~^2P^o_{1/2,3/2}$ of $^{19}$F shows that electron correlation effects reach at maximum 2.2\%, as illustrated by columns 2-4 of table~\ref{ER}. The two last columns reveal that relativistic effects are even smaller, around $\le 1~\%$. 
	It is worth noting that the off-diagonal  hyperfine constant $A_{3/2,1/2}=446(10)$~MHz (not reported in the tables), determined by  Radford \textit{et al.}~\cite{Radetal:61a} from microwave Zeeman spectroscopy,  is in good agreement with our theoretical values of 455~MHz, 459~MHz and 468~MHz corresponding respectively to MR4-MCHF, MR4-BP and MR4-MCDHF calculations.
\begin{table}[H]
\caption{Relative differences  $\Delta(a,b) = (a-b)/a$ (in \%) in the hyperfine constants $A_J$ calculated with approaches $a$ and $b$. See text for the explanations of the different notations used for these approaches. }
\begin{center} 
\begin{tabular}{lccccccccc} 
 \hline \hline \\
 [-0.2cm]
 $A_J$ &  $\Delta\text{(SR-MCHF, HF) }$   &&  $\Delta\text{(MR4-MCHF, SR-MCHF) }$&& $\Delta\text{(MR4-MCHF, HF)}$    &&  $\Delta\text{(SR-BP, SR-MCHF)} $   &&  $\Delta\text{(MR4-BP, MR4-MCHF)} $\\
 [0.2cm]
 \hline \\[-0.2cm]
                                                                                                       \multicolumn{10}{c}{$2p^5 \; ^2P^o_{1/2,3/2}$ }     \\
 [0.2cm]                                         
\hline \\[-0.2cm]
$A_{1/2}$&                 1.68                           &&                   $-$0.76                                   &&                    0.93                            &&                     0.87                                 &&                      0.87                 \\
$A_{3/2}$&              $-$2.19                         &&                     1.85                                      &&                 $-$0.30                          &&                     0.06                                 &&                      0.02                 \\
 [0.2cm] 
 \hline \\[-0.2cm]
                                                                                                         \multicolumn{10}{c}{$2p^4(^3P)3s \; ^2P_{1/2,3/2} $}    \\
  [0.2cm]                                   
\hline    \\[-0.2cm]                                   
$A_{1/2}$ &              $-$21.39                       &&                  $-$3.33                                   &&                 $-$25.44                        &&                     6.71                                 &&                      6.70                 \\
$A_{3/2}$ &               $-$4.49                        &&                     0.08                                     &&                 $-$4.40                          &&                   $-$2.18                                 &&                   $-$2.04               \\
 [0.2cm] 
 \hline\\[-0.2cm]
                                                                                                          \multicolumn{10}{c}{$2p^4(^3P)3s \; ^4P_{1/2,3/2,5/2} $}     \\
 [0.2cm]                                        
\hline \\[-0.2cm] 
 $A_{1/2}$ &           $-$192.52                     &&                   $-$48.00                                   &&                 $-$333.34                        &&                   27.02                                 &&                     35.49                    \\
 $A_{3/2}$ &               24.71                        &&                       7.31                                      &&                    30,21                            &&                     6.08                                 &&                       5.43                       \\
 $A_{5/2}$ &                 3.99                        &&                       0.74                                      &&                     4.70                             &&                     0.01                                 &&                       0.                      \\
  [0.2cm]
  \hline \\[-0.2cm]
                                                                                                              \multicolumn{10}{c}{ $2p^4(^3P)3p \; ^4S^o_{3/2} $ }     \\
     [0.2cm]                                         
\hline \\[-0.2cm]
$A_{3/2}$ &                                               &&                      29.41                                     &&                                                      &&                   146.40                         &&        181.58                            \\

 \hline \hline 
\end{tabular}
\end{center}
\label{ER}
\end{table}

As far as the two levels of $2p^4(^3P)3s \; ^2P_{1/2,3/2}$ are concerned, one can deduce from columns 2-4 that most of the electron correlation is captured through the single reference calculations. These corrections are much larger for $A_{1/2}$ than for $A_{3/2}$. Unlike the ground term, relativistic corrections are quite large (columns 5-6) and cannot be neglected, particularly for $J=1/2$.  \\

\subsubsection{$2p^4(^3P)3s \; ^4P $ and  $2p^4(^3P)3p \; ^4S^o$} 

According to the present results, both states $2p^4(^3P)3s \; ^4P $ and  $2p^4(^3P)3p \; ^4S^o$ have hyperfine structures that appear to be much more sensitive to electron correlation and relativistic effects than the levels considered in the previous section. This is especially true for the $^{4}P_{1/2}$  and  $^{4}S^o_{3/2}$ states.\\

The $2p^4(^3P)3s \; ^4P $ hyperfine constants are given in  Table~\ref{tab:3s4P} while the relative differences between different models are presented  in  Table~\ref{ER}. It can be seen that  HF and SR-MCHF models lead to very different  $A_{1/2}$ values. Moreover, the contribution of electron correlation remains large beyond the SR-MCHF method. It  reaches indeed around 48\% when comparing the MR4-MCHF[10g] and SR-MCHF[10g] results. It would therefore be interesting to go beyond the MR4-MCHF correlation model in order to investigate the hyperfine constants convergence but the calculations become unmanageable due to the large dimensions of the  configuration spaces.
 We therefore adopted another computational strategy to estimate the effect of enlarging the multireference space from MR4 to MR6 and MR8 for this [10g] orbital active set.
We first performed SR and MR$x$-MCHF $(x=3,4)$ calculations for both [8g] and [10g] active sets to estimate the 
$\Delta A_J^{[10g-8g]} (\alpha) = A_J^{[10g]}(\alpha) - A_J^{[8g]} (\alpha)$ differences using the three models ($\alpha =$ SR-MCHF, MR3-MCHF and MR4-MCHF). These differences being very similar (within 0.4~MHz), one added the average difference
\[ \Delta A_J (10g-8g) = 
\frac{1}{3} \sum_\alpha \Delta A_J^{[10g-8g]} (\alpha)  \]
to the MR6-MCHF[8g] and  MR8-MCHF[8g] values. These estimated values are reported with a subscript ``$est.$'' in Table~\ref{tab:3s4P}. 
We estimated the hyperfine constants corresponding to  MR4-BP[10g],  MR6-BP[10g] and  MR8-BP[10g] in a similar way, by reporting the relativistic corrections (MCHF $-$ BP) obtained with SR- and MR3-MCHF models  on the MR$x$-BP~$(x=4, 6$ and $8)$ $A_J$-values.

\begin{table}[H]
\caption{Hyperfine structure constants $A_J$ (in MHz) of $2p^4(^3P)3s \; ^4P $ of $^{19}$F~I. $N_c$ is the number of CSFs. AS specifies the orbital active set (see text for the notations). }
\begin{center}
\begin{tabular}{llcccc}\\ 
\hline \hline \\[-0.2cm]
Method       &     AS       &   $N_c$   &   $A_{1/2}$    &     $A_{3/2}$ &     $A_{5/2}$     \\
[0.2cm]
\hline  \\[-0.2cm]  
\multicolumn{6}{c}{Non Relativistic} \\ 
[0.2cm]                     
\hline     \\[-0.2cm]          
HF                &               &     1         &    $-$908.68      &           305.90 & 2503.96           \\
SR-MCHF   &      [8g]          &  16547    &  $-$305.23         &     412.97     &     2611.21        \\
SR-MCHF   &     [10g]           &  31161    &  $-$310.64         &     406.27     &     2608.12        \\
[0.2cm]
\hline \\[-0.2cm]  
MR3-MCHF  & [8g]  & 123861  &  $-$213.26          &     445.17     &    2633.06         \\
MR3-MCHF  & [10g]& 236829  &  $-$218.76          &     438.12     &    2629.65          \\
MR4-MCHF  & [8g]  & 135265  &  $-$204.31          &     445.39     &     2630.99         \\
MR4-MCHF  & [10g]& 260638  &  $-$209.69          &     438.29     &     2627.49         \\
MR6-MCHF  & [8g]  & 295222  & $-$170.46           &     462.37     &     2646.14          \\ 
MR6-MCHF  &   [10g]$_{est.}$     &        & $-$175.89           &     455.42     &     2642.81          \\
MR8-MCHF  & [8g]  & 331429  & $-$155.53           &     465.92     &     2647.36          \\
MR8-MCHF  &  [10g]$_{est.}$     &          &  $-$160.96          &      458.97    &     2644.03           \\  
[0.2cm]                   
\hline  \\[-0.2cm]  
\multicolumn{6}{c}{Relativistic} \\ 
[0.2cm]                     
\hline        \\[-0.2cm]             
SR-MCDHF   &    [10g]    & 366764  &  $-$502.70          &        404.39    &      2576.51            \\
[0.2cm] 
\hline \\[-0.2cm]  
SR-RCI-P          &    [10g]    &  716054 &  $-$441.08         &        426.60    &     2599.17              \\
MR3-RCI-P &  [10g,10g,6g]             &1095434 &    $-$356.42        &        455.36     &     2618.79              \\ 
[0.2cm]
\hline \\[-0.2cm]  
SR-BP         &    [10g]   &  370941 & $-$425.65          &        432.22    &     2608.39             \\
MR3-BP      &    [10g]   & 594890  &  $-$334.46          &        463.30    &     2629.26            \\
MR4-BP      &    [10g]$_{est.}$            &              &   $-$325.03           &     463.86       &      2627.43             \\
MR6-BP      &    [10g]$_{est.}$            &              &   $-$291.23           &    480.99        &      2642.75            \\
MR8-BP      &    [10g]$_{est.}$            &              &   $-$276.30          &     484.54       &    2643.97               \\
[0.2cm]  
\hline  \\ [-0.2cm]  
\multicolumn{3}{l}{Observed} &                   &                         &        \\
\multicolumn{3}{l}{Levy \etal~\cite{Levetal:07a}} &                   &                         &  $2643 \pm 1$       \\
[0.2cm]
\hline \hline \\
\end{tabular}
\end{center}
\label{tab:3s4P}
\end{table}

As can be seen in Table~\ref{ER}, the contribution of the relativistic effects on the constant $A_{1/2}$ reaches 35\% when going from MR4-MCHF to MR4-BP. The $A_{3/2}$ constant is also sensitive to electron correlation and, to a lesser degree (6\%), to relativistic corrections. On the contrary, correlation effects are rather small and relativity is negligible for the constant  $A_{5/2}$. All these observations are confirmed by SR-MCDHF, SR-RCI-P and MR3-RCI-P calculations that also illustrate a satisfactory consistency between independent SR-RCI-P/SR-BP, and MR3-RCI-P/MR3-BP calculations.  
Our recommended values for $A_{1/2}$, $A_{3/2}$ and $A_{5/2}$ are respectively $-$276.30~MHz,  484.54~MHz, 2643.97~MHz corresponding to the MR8-BP estimation. To the knowledge of the authors, there are no experimental values available for the first two constants. For $ A_ {5/2} $, our theoretical value is in excellent agreement with the hyperfine structure constant deduced from   laser-induced fluorescence and modulated optical depopulation pumping experiment~\cite{Levetal:07a}. The relative uncertainty, previously defined, for this constant is less than 0.1\%. 
\\

Table~\ref{tab:3p4S} reports the calculations of the hyperfine constant $A_{3/2}$ for the $2p^4(^3P)3p \; ^4S^o$ level in various multiconfigurational approximations. There is no experimental value available for comparison. 
The three contributions,  $A_{orb}$, $A_{sd}$ and $A_c$ of equations (\ref{Aorb}), (\ref{Asd}) and (\ref{Ac}) are strictly zero in the single configuration Hartree-Fock approximation. The $J$-independent hyperfine parameters defined by equations  (\ref{al}), (\ref{asd}) and (\ref{ac}) constitute the basic ingredients of the hyperfine constants. The first two, $a_l$ and $a_{sd}$, both contain a  $\sqrt{L}$ factor that annihilates the $A_{orb}$ and $A_{sd}$ contributions for $^4S^o$. The third contribution, due to the contact contribution may resist for $S \ne 0$ but cancels in the single-configuration approximation because there is no open $s$-shells in the configuration considered.
The hyperfine $a_l$ and $a_{sd}$  parameters  remain zero beyond the single configuration approximation (in the SR-MCHF, MR3-MCHF and MR4-MCHF correlation models) and considering that $L$ and $S$ are good quantum numbers, i.e. omitting any term-mixing due to relativity. Oppositely, the $a_c$ parameter becomes different from zero when  spin-polarisation contributions involving $1s, 2s \rightarrow s'$ single-electron excitations of the core closed shells, or configurations involving open $ns$-subshells are included in the wave function expansions~\cite{Godetal:97b}. One can read in Table~\ref{tab:3p4S}  that the null value of $A_c$ switches to 95~MHz in the SR-MCHF model and further increases to 132~MHz when extending the multireference space (MR4-MCHF). 
\begin{table}[H]
\caption{Hyperfine structure constant $A_{3/2}$ (in MHz) of $2p^4(^3P)3p \; ^4S^o $ of $^{19}$F~I, with the three  $A_{orb}$, $A_{sd}$ and $A_c$ contributions (see Eqs.~\ref{Aorb}-\ref{Ac} for their definition). See text for the various notations used for the theoretical methods. $N_c$ is the number of CSFs. AS specifies the orbital active set.}
\begin{center}
\begin{tabular}{llccrrrr}\\ 
\hline \hline \\[-0.2cm]

Method       &     AS       &   $N_c$   &  $A_{orb}$  &  $A_{sd}$  &    $A_c$       &   $A_{3/2}$    \\
[0.2cm]
\hline  \\[-0.2cm]  
                   \multicolumn{7}{c}{Non Relativistic} \\ 
[0.2cm]                     
\hline     \\[-0.2cm]          
HF                &               &     1         &      0.            &       0.            &          0.            &    0.           \\
SR-MCHF   &     [10g]   &  27119    &       0.            &      0.             &       95.52         &  95.52               \\
[0.2cm]
\hline \\[-0.2cm]  
MR3-MCHF  &   [10g]    & 137533 &       0.            &          0.          &        135.32       &  135.32                  \\
MR4-MCHF  &   [10g]    & 211516 &       0.            &          0.          &       132.68        &  132.68              \\
[0.2cm]                   
\hline  \\[-0.2cm]  
                \multicolumn{7}{c}{Relativistic} \\ 
[0.2cm]                     
\hline        \\[-0.2cm]   
($2p^4 3p + 2p^5$)-MCDHF   &     & 9&&&               &  $-$358.64           \\          
SR-MCDHF   & [10d7f6g]     & 226288 &&&               &  $-$248.64           \\
MR3-RCI   &  [10d7f6g,5g,5g]    &1288207&&&               &       $-$156.87    \\
[0.2cm]  
\hline \\[-0.2cm] 
SR-RCI-P   & [10g] & 621284  &&&               & $-$206.32            \\
SR-RCI-P*& [10g]    &621284 &&&               &  $-$211.52          \\
[0.2cm]  
\hline \\[-0.2cm] 
($2p^4 3p + 2p^5$)-BP         &                 &      9      &  $-$401.12&   22.68      &     0.                  &  $-$378.44  \\
($2p^4 3p + 2p^5$)-BP ($^4P^o$ excluded) &                 &      8       &  0.56 &   $-$11.21   &     0.                  &  $-$10.64 \\
SR-BP         &    [10g]    &  621272 &    $-$307.38 &  6.61            &     94.93            &  $-$205.84          \\
MR3-BP      &   [10g,5g,5g]              &  816353 &   $-$296.86  &     3.69         &    132.61           &  $-$160.56          \\
MR4-BP      &   [10g,5g,5g,5g]  &  1011587& $-$296.19 &     3.53         &    130.02            &   $-$162.64           \\
[0.2cm]  
\hline \hline \\
\end{tabular}
\end{center}
\label{tab:3p4S}
\end{table}

The second part of Table~\ref{tab:3p4S} illustrates the huge relativistic effects, as quantified in the last line of Table~\ref{ER}. The relative variation between MR4-BP and MR4-MCHF reaches indeed $\Delta\% = 182\%$. A first relativistic approach based on the $(2p^4 3p + 2p^5) (3/2)^o$ MCDHF calculation targeting the fifth eigenvector of $J=(3/2)^o$ symmetry, gives a quite large but negative $A_{3/2}$ value, in contradiction with the multireference MCHF results. Electron correlation reduces the absolute value of the hyperfine constant by more than a factor two but the MR3-RCI negative value remains significantly large.
These a priori unexpected relativistic effects can be explained by investigating the content of the hyperfine constant in the Breit-Pauli approximation.
By first performing the Breit-Pauli ($2p^4 3p + 2p^5$) calculation, including all the $J=3/2$ lower states from $2p^4 3p$ and $2p^5$, one realized that a large negative $A_{orb}$ contribution appears due to the relativistic mixing with the $^{4}P^o$ configuration states. This is confirmed by limited BP calculations excluding this symmetry block, as illustrated in Table~\ref{tab:3p4S}. The 20~MHz agreement between the ($2p^4 3p + 2p^5$)  BP results and the corresponding Dirac-Hartree-Fock values of $A_{3/2}$ for which the three magnetic dipolar interactions are not separable, is very satisfactory, considering the huge effect found for this hyperfine constant. 
Beyond these limited correlation models, one observes a satisfactory agreement between SR-MCDHF, SR-BP and SR-RCI-P results for the single-reference approaches, and between MR3-RCI and MR3-BP values for the multireference calculations. Note that the [10d7f6g] notation for the orbital active set (AS)  used in SR-MCDHF calculation implies the limitation $l_{\mbox{max}}=2$  for $n \leq 10$, $=3$ for $n \leq 7$, $=4$ for $n \leq 6$. The QED effects are estimated being a few MHz by comparing the SR-RCI-P* results that neglects QED corrections by setting $g_s = 2$ in eq.~(\ref{Asd}) and (\ref{Ac}) with the SR-RCI-P values that include them.

\subsubsection{Analysis of  differential relativistic effects}

 In order to investigate the differential relativistic effects on the hyperfine constants for the different $J$-values of a given term, we report in  Table~\ref{tab:A_i_contribution}, for all states considered, the three contributions $A_{orb}$,  $A_{sd}$ and $A_{c}$ to the hyperfine constant as well as their relative variations due to the inclusion of relativistic corrections in the Breit-Pauli approximation, defined by:
$$\frac{\Delta A_{i}}{A_J} = \frac{A_{i}(\text{MR4-BP}) - A_{i}(\text{MR4-MCHF})}{A_J(\text{MR4-BP})} \;\;\;\; \mbox{with} \;\;i=orb, ~ sd,~ c.$$  
\begin{table}[H]
\caption{Details of the different  contributions (all in MHz) to the hyperfine constant $A_J$ : orbital ($A_{orb}$) , spin-dipolar ($A_{sd}$) and  contact ($A_{c}$) as well as their relative variations  $\Delta A_{i}/A_J$  due to  the inclusion of  relativistic corrections (see text for explanations).}
\begin{center}
\begin{tabular}{llrrrrrrrrr} \\
 \\
\hline \hline \\[-0.2cm]
 &                                                  &                 \multicolumn{3}{c}{$J=1/2$ }                         &                          \multicolumn{3}{c}{$J=3/2$ }        &    \multicolumn{3}{c}{$J=5/2$} \\  [0.1cm]
\hline  \\[-0.2cm]
                               &                    &     MR4-MCHF &    MR4-BP        &  $\Delta A_{i}/A_J$ & MR4-MCHF   &    MR4-BP   &  $\Delta A_{i}/A_J$ & MR4-MCHF &    MR4-BP   &  $\Delta A_{i}/A_J$   \\  [0.1cm]
\hline  \\[-0.2cm]
$2p^{5}~^{2}P^o$  &  $A_{orb}$  &      4905.88    & 4948.86       &          0.42~\%            &   2452.94    &  2453.66 &   0~\%                         &                  &                  &                                        \\
                              &   $A_{sd}$   &      5386.16    &  5428.31      &          0.41~\%             &   $-$538.62  & $-$538.44&    0~\%                        &                  &                 &                                        \\
                              &   $A_{c}$     &      $-$97.75   &  $-$93.43       &          0.04~\%             &   97.75        &     97.16  &    0~\%                        &                  &                 &                                        \\ 
 &&&&&&&&&& \\   
                             &    $A_J$   &    10194.28    & 10283.74       &       0.87~\%                 &  2012.07    & 2012.38  &      0~\%                       &                  &                 &                                         \\
                             &                     &                        &                        &                                     &                     &                &                                      &                  &                 &                                      \\
&&&&&&&&&& \\ 
$3s~^{2}P$          &   $A_{orb}$   &   5610.41        &   5611.23        &    0.05~\%                          &   2805.20   & 2802.40   &  $-$0.09~\%              &                 &                 &                                          \\
                            &   $A_{sd}$     & $-$4067.54     & $-$4032.05    &     2~\%                           &   406.75      & 296.24     &  $-$3.64~\%         &                 &                 &                                          \\
                            &  $A_{c}$        & 113.06            &  195.68          &       4.66~\%                  &   $-$113.06  &$-$61.74    &  ~1.69~\%              &                   &                &                                          \\
 &&&&&&&&&& \\           
                            &   $A_J$    & 1655,93          &  1774,.85       &    6.71~\%                    & 3098.89     &   3036.90   &   $-$2.04~\%               &                  &                  &                                          \\ 
    &&&&&&&&&& \\      
$3s~^{4}P$          &  $A_{orb}$    &  $-$2815.60    &$-$2829.30   &   4.21~\%                    & 1126.24     & 1124.49   &  $-$0.46~\%                & 1689.36    & 1691.64  &    0.09~\%                        \\
                            &   $A_{sd}$     &  1019.20        & 975.10          &   13.57~\%                  &$-$1386.11 &$-$1317.02& 14.89~\%                      &   366.91      & 354.89    &$-$0.46~\%     \\
                             &$A_{c}$         &   1586.72       & 1529.13       &  17.72~\%                       & 698.16     &  656.39  &     $-$9.00~\%             &   571.22       &  580.90&  0.37~\%  \\
   &&&&&&&&&&\\ 
                            &      $A_J$  &  $-$209.69     &  $-$325.03 &   35.49~\%                     & 438.29       &   463.86     &      5.89~\%              & 2627.49      & 2627.43 & 0~\%     \\
   &&&&&&&&&&\\ 
  $3p~^{4}S$       &   $A_{orb}$    &                        &                          &                                       & 0.                 & $-$296.19   & 182.13~\%        &                  &                 &                                      \\  
                            &  $A_{sd}$      &                        &                          &                                       & 0.                 & 3.52       & $-$2.16~\%            &                  &                 &                                       \\  
                            & $A_{c}$         &                        &                          &                                       & 132.68         & 130.02        &  1.64~\%          &                  &                 &                                      \\    
 &&&&&&&&&&\\ 
                            &  $A_J$      &                        &                          &                                       & 132.68        &$-$162.64     & 181.58~\%      &                  &                 &                                      \\               
    &&&&&&&&&& \\ [0.1cm]
\hline \hline \\
\end{tabular}
\end{center}
\label{tab:A_i_contribution}
\end{table}
Whereas relativistic effects are found to be very low  on the ground state hyperfine structure, for all three kinds of magnetic dipole hyperfine interactions, they are not negligible for $A_{sd}$ and $A_{c}$ of the $2p^4(^3P)3s \; ^2P_{1/2,3/2}$ states. These corrections having the same signs for $ A_ {1/2} $ but different signs for $ A_ {3/2} $, the global effect reaches 7\% for $J=1/2$ but reduces to 2\% for $J=3/2$.
A more spectacular effect is found for the spin-dipolar and contact contributions of $A_{1/2}$  and  $A_{3/2}$ of the $2p^4(^3P)3s \; ^4P $. The relative variations are, respectively,  of the order of 14\%  and 18\% for $A_{1/2}$ and 15\%  and 9\% for $A_{3/2}$. The two effects are cumulative for $A_{1/2}$ while they cancel each other for $A_{3/2}$, largely reducing the global effect for $J=3/2$.  The $2p^4(^3P)3p \; ^4S^o $ state is certainly the most remarkable case. The relative variation due to relativity exceeds indeed 180\%. As already discussed in the previous section, this effect is mostly due to 
$^4S^o - {^4P}^o$ term-mixing that directly affects the orbital $ A_{orb} $ contribution.\\

\subsection{$2p^3(^4S)3s \; ^5S^o_{2}$ and  $2p^3(^4S)3p \; ^5P_{1,2,3}$ of $^{17}$O~I.}

We revisited the \textit{ab-initio} calculation of the hyperfine constants of   two excited states $2p^3(^4S)3s \; ^5S^o$ and  $2p^3(^4S)3p \; ^5P$ of $^{17}$O~I that were previously evaluated using the nonrelativistic multiconfigurational Hartree-Fock method~\cite{Godetal:97b,JonGod:00a}. Large discrepancies were found with observation~\cite{Maretal:93a,JenTat:2000a}, reaching more than 20\% in some cases. In the present work, we revisit the theoretical evaluation of these hyperfine constants, investigating the relative importance of correlation  and relativistic effects, as for the fluorine states. These constants have been estimated using $I=5/2$,  $\mu = -1.89379$~nm and 
$Q = -0.02579$~barn, taken from Stone's compilation~\cite{Sto:2005a}.
This $Q(^{17}$O) value falls within the error limits reported in Pyykk\"o's compilation~\cite{Pyy:2008a}, $Q = -0.02558(22)$~barn, and adopted in Stone's 2016 most recent compilation of nuclear electric quadrupole moments~\cite{Sto:2016a}. 
\\

Table~\ref{tab:A_J_5S} presents the magnetic dipole constant $A_2$ for the state $2p^3(^4S)3s \; ^5S^o_{2}$. 
In the same table we also report the $a_c$ hyperfine parameter that is the only one making $A_2$ in nonrelativistic calculations,   $a_{sd}$ and  $a_l$ being strictly zero because $L=0$.  
\begin{table}[H]
\caption{Hyperfine structure parameter $a_c$ (in a.u.) and constant $A_J$ (in MHz) of  $2p^3(^4S)3s \; ^5S^o_{2}$ in O~I.  $N_c$ is the number of CSFs. AS specifies the orbital active set (see text for the notations). }
\begin{center} 
\begin{tabular}{lccccccc} 
 \\
\hline \hline \\[-0.2cm]
Method                               &&     AS   & &      $N_c$     & & $a_c$ &            $A_{2}$                  \\ [0.1cm]
\hline  \\[-0.2cm]
     HF                                 & &              & &         1          &  &  &    $-$63.62                            \\
[0.2cm]
  SR-MCHF                        & &    [10g]  &&       8701        & & 7.3981   &$-$89.22                             \\
  [0.2cm]
MR3-MCHF                       & &               &&      38413       & & 7.8812 &   $-$95.05                             \\
 MR4-MCHF                      & &               &&      44855       &  & 7.8675 &   $-$94.88                               \\         
 MR6-MCHF                      &  &              &&      61278       &  & 7.9532  &  $-$95.91                                 \\  
 [0.2cm]                            
MR6~$\cup$~TQ-CI                       & &              &&      661468     &  & 8.1128  &   $-$97.84                                 \\
[0.2cm]
\hline  \\
 SR-BP                              & &     [10g]   &&     303286     &  &  &  -89.71                                    \\      
 [0.2cm]
\hline  \\[-0.2cm]
This work                        &  &                & &                    &  &    &                 $-$98.33                 \\
\hline 
\\[-0.2cm]  \multicolumn{2}{l}{Other theories} &&                       &&&&  \\
\\[-0.2cm]
\multicolumn{2}{l}{Godefroid \etal~\cite{Godetal:97b}}     &&                       &&&&           $-$96.70                                  \\
\multicolumn{2}{l}{J\"onsson and Godefroid~\cite{JonGod:00a}}     &&                       &&& &          $-$91.95              \\ [0.1cm]
\hline
\\[-0.2cm]  \multicolumn{2}{l}{Observation} &&                       &&&&  \\  
\\[-0.2cm]
  \multicolumn{2}{l}{Marin \etal~\cite{Maretal:93a} }    &&                       &&& &   $ -98.59(43) $       \\
\multicolumn{2}{l}{Jennerich and Tate~\cite{JenTat:2000a}}       &&              && &&    $ -97.93 (10)$           \\[0.1cm]
\hline \hline \\
\end{tabular}
\end{center}
\label{tab:A_J_5S}
\end{table}
\clearpage
On can deduce from the  $\Delta$(SR-BP, SR-MCHF) relative differences ($\simeq 0.5\%$) that relativistic effects on this hyperfine constant are quite small.
We therefore focused our theoretical approach on the description of correlation effects. For this, we performed MR$x$-MCHF calculations, where $x=3, \; 4, \; 6$ refer to the weight analysis presented in  Table~\ref{tab:weights}. The convergence of the hyperfine parameter with respect of the MR extension is around 1\%. Higher correlation effects beyond the SD-MR-MCHF model were included through  a configuration  interaction calculation combining the MR6-MCHF expansion with a configuration subspace including triple and quadruple excitations from the main reference $1s^{2}2s^{2}2p^{3}3s$ up to [6f]. This calculation is labeled MR6 $\cup$ TQ-CI. 
The values of $ a_c $ and $A_2$ vary by only 2\% when going from  SR-MCHF  to  MR6 $\cup$ TQ-CI. We can then consider that convergence is achieved within a few percents. If we add the difference $A_2 (\text{SR-BP}) - A_2 (\text{SR-MCHF}) - $ representing the relativistic correction, to the corresponding to the MR6 $\cup$ TQ-CI value,  we obtain  $A_2 = -98.33$~MHz. We consider this latter value as our recommended value, in very good agreement with experiments  \cite{Maretal:93a,JenTat:2000a}.  The relative uncertainty is around 0.4\%.\\

The magnetic and quadrupole electric hyperfine constants of $ 2p ^ 3 (^ 4S) 3p \; ^ 5P_ {1,2,3} $ are reported in  Table~\ref{tab:A_J_5P}. 
\begin{table}[H]
\caption{Hyperfine structure constants $A_J$ and $B_J$ (in MHz) of   $2p^3(^4S)3p \; ^5P_{1,2,3}$ of $^{17}$O~I. $N_c$ is the number of CSFs. AS specifies the orbital active set (see text for the notations).}
\begin{center} 
\begin{tabular}{lcccccccc} 
 \\
\hline \hline \\[-0.2cm]
Method                 &     AS                &   $N_c$  &      $A_{1}$  &   $A_{2}$   &   $A_{3}$          &$B_{1}$ &$B_{2}$ &  $B_{3}$         \\ [0.1cm]
\hline  \\[-0.2cm]
      HF                   &                            &        1    &         5.40     &     $-$3.18     &   $-$2.21               & $-$0.0278               &  0.2785       &   $-$0.2785               \\
[0.2cm]   
 SR-MCHF           &     [10g]             &    46826   &      $-$15.16    &     $-$17.97    &      $-$13.60         & $-$0.0439   &    0.4394    &  $-$0.4394   \\
[0.2cm]                                
MR3-MCHF          &     [10g]                     &   215838  &     $-$27.24    &    $-$24.81     &    $-$19.09            &  $-$0.0447  &    0.4465    &  $-$0.4465      \\
MR4-MCHF          &    [10g]                        &  241908  &    $-$27.93     &    $-$25.17   &     $-$19.38          &   $-$0.0447   &    0.4468   &  $-$0.4468       \\         
MR6-MCHF          &    [10g]                        & 298454  &      $-$26.84   &     $-$24.60    &       $-$18.92        & $-$0.0442   & 0.4422      &   $-$0.4422                   \\ 
[0.2cm] 
\hline  \\[-0.2cm]                                                          
 SR-BP                &      [10g]              & 845623  &     $-$13.84     &    $-$17.40    &   $-$13.04          &   $-$0.0441   &    0.4406   &  $-$0.4399       \\ 
MR3-BP              &      [10g]               & 1336911 &     $-$25.52    &     $-$24.01  &    $-$18.34          &   $-$0.0447   &    0.4470   &  $-$0.4462       \\ 
[0.2cm]
\hline  \\[-0.2cm]
\multicolumn{2}{l}{Observation:}   & &&&&&&  
\\                                                     
\multicolumn{2}{l}{Jennerich and Tate~\cite{JenTat:2000a}}    &              &     $-25.83 (10)$          & $-24.47 (11)$ &  $-18.70 (5)$  &$0.0(2)$&$0.9(8)$&$ -0.2(5) $\\
\multicolumn{2}{l}{Marin \etal~\cite{Maretal:93a}}  &               &    $-$26.39 &                   &                            &                                  &                    &                           \\
[0.2cm]
\hline  \\[-0.2cm]
\multicolumn{2}{l}{Other theory:}   & &&&&&&  
\\    
 \multicolumn{2}{l}{J\"onsson and Godefroid~\cite{JonGod:00a}} &               &    $-$18.73    &        $-$19.88  &          $-$15.16      &                &                    &                 \\ [0.1cm]
\hline  \\[-0.2cm]
 \hline \hline \\    
\end{tabular}
\end{center}
\label{tab:A_J_5P}
\end{table}
Our results show that the electric quadrupole interaction is  smaller  than the magnetic dipolar interaction, as found by Jennerich and Tate~\cite{JenTat:2000a}.  The relative variations $\Delta$(SR-BP, SR-MCHF) for $A_1$, $A_2$ and $A_3$ of   $2p^3(^4S)3p \; ^5P$ are, respectively, 10\%, 3\% and 4\%. The effect of relativity on the hyperfine structures of this term cannot therefore be neglected. From the evolution of the hyperfine parameters when extending the MR space, one can conclude that electron correlation is well described with the MR3-MCHF approach.  We therefore performed the MR3-BP calculation with a  configuration space of 1~336~911 CSFs. The results obtained are in very good agreement with observation. The relative uncertainties are of the order of 1\% for $A_1$ constant and 2\% for  $A_2$ and $A_3$. 
 
\section{Conclusion}

We calculated the hyperfine constants of  $2p^5 \; ^2P^o$, $2p^4(^3P)3s \; ^2P$, $2p^4(^3P)3s \; ^4P $ and  $2p^4(^3P)3p \; ^4S^o$ of $^{19}$F~I and $2p^3(^4S)3s \; ^5S^o$ and  $2p^3(^4S)3p \; ^5P$ of $^{17} $O~I using various multiconfigurational correlation models. Large effects of electron correlation and relativity are found on the hyperfine structure of some atomic states of fluorine and oxygen. This is particularly the case  of  $A_{3/2}$   ($2p^4(^3P)3p \; ^4S^o $ ),  $A_{1/2}$ ($2p^4(^3P)3p \; ^4P $)  of F~I and $A_{1}$ ($2p^3(^4S)3p \; ^5P$) of O~I. For all states, the inclusion of higher-order correlation effects through the use of multireference spaces was found to be necessary. In the case where relativistic effects are important, the three relativistic approaches BP, RCI-P and MCDHF are consistent with each other. The excellent agreement between theory and available experimental values from Doppler-free saturated absorption spectra indicate that the analysis of the latter is correctly done with no problems to interpret the crossover signals.

The analysis of the different contributions to the magnetic dipole interaction in the Breit-Pauli approximation remains a very  useful tool to unravel their variations with correlation and relativity. Light atomic systems of the second period remain interesting probes for benchmarking the theoretical models. With this respect we hope that the present work will stimulate further experimental measurements, in particular
 involving the $2p^4(^3P)3p \; ^4S^o_{3/2} $ and $2p^4(^3P)3s \; ^4P_{1/2,3/2} $of $^{19}$F~I.
 The hyperfine structures of different atomic transitions  will be measured in $^{17,18,19}$F by collinear laser spectroscopy, with the ultimate goal of determining the charge radii of exotic fluorine isotopes~\cite{Ruietal:2016a}. In that framework, the present calculations are valuable to guide the planned experiments of exotic fluorine isotopes. \\

\vspace{-0.5cm}

\begin{acknowledgments}
N.A. and M.N. acknowledge financial support from the Direction G\'en\'erale de la Recherche Scientifique et du D\'evelopement Technologique (DGRSDT) of Algeria. P.J. acknowledges financial support from the Swedish Research Council (VR), under contract 2015-04842. This work has also been partially supported by the Belgian F.R.S.-FNRS Fonds de la Recherche Scientifique (CDR J.0047.16) and the BriX IAP Research Program No. P7/12 (Belgium). Computational resources have been partially provided by the Consortium des Équipements de Calcul Intensif (CÉCI), funded by the Fonds de la Recherche Scientifique de Belgique (F.R.S.-FNRS) under Grant No. 2.5020.11.

\end{acknowledgments}

\begin{thebibliography}{30}%
\makeatletter
\providecommand \@ifxundefined [1]{%
 \@ifx{#1\undefined}
}%
\providecommand \@ifnum [1]{%
 \ifnum #1\expandafter \@firstoftwo
 \else \expandafter \@secondoftwo
 \fi
}%
\providecommand \@ifx [1]{%
 \ifx #1\expandafter \@firstoftwo
 \else \expandafter \@secondoftwo
 \fi
}%
\providecommand \natexlab [1]{#1}%
\providecommand \enquote  [1]{``#1''}%
\providecommand \bibnamefont  [1]{#1}%
\providecommand \bibfnamefont [1]{#1}%
\providecommand \citenamefont [1]{#1}%
\providecommand \href@noop [0]{\@secondoftwo}%
\providecommand \href [0]{\begingroup \@sanitize@url \@href}%
\providecommand \@href[1]{\@@startlink{#1}\@@href}%
\providecommand \@@href[1]{\endgroup#1\@@endlink}%
\providecommand \@sanitize@url [0]{\catcode `\\12\catcode `\$12\catcode
  `\&12\catcode `\#12\catcode `\^12\catcode `\_12\catcode `\%12\relax}%
\providecommand \@@startlink[1]{}%
\providecommand \@@endlink[0]{}%
\providecommand \url  [0]{\begingroup\@sanitize@url \@url }%
\providecommand \@url [1]{\endgroup\@href {#1}{\urlprefix }}%
\providecommand \urlprefix  [0]{URL }%
\providecommand \Eprint [0]{\href }%
\@ifxundefined \urlstyle {%
  \providecommand \doi  [0]{\begingroup \@sanitize@url \@doi}%
  \providecommand \@doi [1]{\endgroup \@@startlink {\doibase
  #1}doi:\discretionary {}{}{}#1\@@endlink }%
}{%
  \providecommand \doi  [0]{doi:\discretionary{}{}{}\begingroup
  \urlstyle{rm}\Url }%
}%
\providecommand \doibase [0]{http://dx.doi.org/}%
\providecommand \Doi [0]{\begingroup \@sanitize@url \@Doi }%
\providecommand \@Doi  [1]{\endgroup\@@startlink{\doibase#1}\@@Doi}%
\providecommand \@@Doi [1]{#1\@@endlink}%
\providecommand \selectlanguage [0]{\@gobble}%
\providecommand \bibinfo  [0]{\@secondoftwo}%
\providecommand \bibfield  [0]{\@secondoftwo}%
\providecommand \translation [1]{[#1]}%
\providecommand \BibitemOpen [0]{}%
\providecommand \bibitemStop [0]{}%
\providecommand \bibitemNoStop [0]{.\EOS\space}%
\providecommand \EOS [0]{\spacefactor3000\relax}%
\providecommand \BibitemShut  [1]{\csname bibitem#1\endcsname}%
\bibitem [{\citenamefont {J\"{o}nsson}\ \emph {et~al.}(2010)\citenamefont
  {J\"{o}nsson}, \citenamefont {Carette}, \citenamefont {Nemouchi},\ and\
  \citenamefont {Godefroid}}]{Jonetal:2010a}%
  \BibitemOpen
  \bibfield  {author} {\bibinfo {author} {\bibfnamefont {P.}~\bibnamefont
  {J\"{o}nsson}}, \bibinfo {author} {\bibfnamefont {T.}~\bibnamefont
  {Carette}}, \bibinfo {author} {\bibfnamefont {M.}~\bibnamefont {Nemouchi}}, \
  and\ \bibinfo {author} {\bibfnamefont {M.}~\bibnamefont {Godefroid}},\
  }\href@noop {} {\bibfield  {journal} {\bibinfo  {journal} {J. Phys. B: At.
  Mol. Opt. Phys.},\ }\textbf {\bibinfo {volume} {43}},\ \bibinfo {pages}
  {115006} (\bibinfo {year} {2010})}\BibitemShut {NoStop}%
\bibitem [{\citenamefont {Jennerich}\ \emph {et~al.}(2006)\citenamefont
  {Jennerich}, \citenamefont {Keiser},\ and\ \citenamefont
  {Tate}}]{Jenetal:06a}%
  \BibitemOpen
  \bibfield  {author} {\bibinfo {author} {\bibfnamefont {R.}~\bibnamefont
  {Jennerich}}, \bibinfo {author} {\bibfnamefont {A.}~\bibnamefont {Keiser}}, \
  and\ \bibinfo {author} {\bibfnamefont {D.}~\bibnamefont {Tate}},\ }\href@noop
  {} {\bibfield  {journal} {\bibinfo  {journal} {Eur. Phys. J. D},\ }\textbf
  {\bibinfo {volume} {40}},\ \bibinfo {pages} {81} (\bibinfo {year}
  {2006})}\BibitemShut {NoStop}%
\bibitem [{\citenamefont {Carette}\ \emph {et~al.}(2010)\citenamefont
  {Carette}, \citenamefont {Nemouchi}, \citenamefont {J\"{o}nsson},\ and\
  \citenamefont {Godefroid}}]{Caretal:10b}%
  \BibitemOpen
  \bibfield  {author} {\bibinfo {author} {\bibfnamefont {T.}~\bibnamefont
  {Carette}}, \bibinfo {author} {\bibfnamefont {M.}~\bibnamefont {Nemouchi}},
  \bibinfo {author} {\bibfnamefont {P.}~\bibnamefont {J\"{o}nsson}}, \ and\
  \bibinfo {author} {\bibfnamefont {M.}~\bibnamefont {Godefroid}},\ }\href@noop
  {} {\bibfield  {journal} {\bibinfo  {journal} {Eur. Phys. J. D},\ }\textbf
  {\bibinfo {volume} {60}},\ \bibinfo {pages} {231} (\bibinfo {year}
  {2010})}\BibitemShut {NoStop}%
\bibitem [{\citenamefont {Krins}\ \emph {et~al.}(2009)\citenamefont {Krins},
  \citenamefont {Oppel}, \citenamefont {Huet}, \citenamefont {von Zanthier},\
  and\ \citenamefont {Bastin}}]{Krietal:09a}%
  \BibitemOpen
  \bibfield  {author} {\bibinfo {author} {\bibfnamefont {S.}~\bibnamefont
  {Krins}}, \bibinfo {author} {\bibfnamefont {S.}~\bibnamefont {Oppel}},
  \bibinfo {author} {\bibfnamefont {N.}~\bibnamefont {Huet}}, \bibinfo {author}
  {\bibfnamefont {J.}~\bibnamefont {von Zanthier}}, \ and\ \bibinfo {author}
  {\bibfnamefont {T.}~\bibnamefont {Bastin}},\ }\href@noop {} {\bibfield
  {journal} {\bibinfo  {journal} {Phys. Rev. A},\ }\textbf {\bibinfo {volume}
  {80}},\ \bibinfo {pages} {062508} (\bibinfo {year} {2009})}\BibitemShut
  {NoStop}%
\bibitem [{\citenamefont {Jennerich}\ and\ \citenamefont
  {Tate}(2000)}]{JenTat:2000a}%
  \BibitemOpen
  \bibfield  {author} {\bibinfo {author} {\bibfnamefont {R.~M.}\ \bibnamefont
  {Jennerich}}\ and\ \bibinfo {author} {\bibfnamefont {D.~A.}\ \bibnamefont
  {Tate}},\ }\Doi {10.1103/PhysRevA.62.042506} {\bibfield  {journal} {\bibinfo
  {journal} {Phys. Rev. A},\ }\textbf {\bibinfo {volume} {62}},\ \bibinfo
  {pages} {042506} (\bibinfo {year} {2000})}\BibitemShut {NoStop}%
\bibitem [{\citenamefont {Tate}\ and\ \citenamefont
  {Aturaliye}(1997)}]{TatAtu:97a}%
  \BibitemOpen
  \bibfield  {author} {\bibinfo {author} {\bibfnamefont {D.~A.}\ \bibnamefont
  {Tate}}\ and\ \bibinfo {author} {\bibfnamefont {D.~N.}\ \bibnamefont
  {Aturaliye}},\ }\Doi {10.1103/PhysRevA.56.1844} {\bibfield  {journal}
  {\bibinfo  {journal} {Phys. Rev. A},\ }\textbf {\bibinfo {volume} {56}},\
  \bibinfo {pages} {1844} (\bibinfo {year} {1997})}\BibitemShut {NoStop}%
\bibitem [{\citenamefont {Carette}\ \emph {et~al.}(2013)\citenamefont
  {Carette}, \citenamefont {Nemouchi}, \citenamefont {Li},\ and\ \citenamefont
  {Godefroid}}]{Caretal:13a}%
  \BibitemOpen
  \bibfield  {author} {\bibinfo {author} {\bibfnamefont {T.}~\bibnamefont
  {Carette}}, \bibinfo {author} {\bibfnamefont {M.}~\bibnamefont {Nemouchi}},
  \bibinfo {author} {\bibfnamefont {J.}~\bibnamefont {Li}}, \ and\ \bibinfo
  {author} {\bibfnamefont {M.}~\bibnamefont {Godefroid}},\ }\Doi
  {10.1103/PhysRevA.88.042501} {\bibfield  {journal} {\bibinfo  {journal}
  {Phys. Rev. A},\ }\textbf {\bibinfo {volume} {88}},\ \bibinfo {pages}
  {042501} (\bibinfo {year} {2013})}\BibitemShut {NoStop}%
\bibitem [{\citenamefont {Carette}\ and\ \citenamefont
  {Godefroid}(2011)}]{CarGod:2011a}%
  \BibitemOpen
  \bibfield  {author} {\bibinfo {author} {\bibfnamefont {T.}~\bibnamefont
  {Carette}}\ and\ \bibinfo {author} {\bibfnamefont {M.~R.}\ \bibnamefont
  {Godefroid}},\ }\Doi {10.1103/PhysRevA.83.062505} {\bibfield  {journal}
  {\bibinfo  {journal} {Phys. Rev. A},\ }\textbf {\bibinfo {volume} {83}},\
  \bibinfo {pages} {062505} (\bibinfo {year} {2011})}\BibitemShut {NoStop}%
\bibitem [{\citenamefont {Radford}\ \emph {et~al.}(1961)\citenamefont
  {Radford}, \citenamefont {Hughes},\ and\ \citenamefont
  {Beltran-Lopez}}]{Radetal:61a}%
  \BibitemOpen
  \bibfield  {author} {\bibinfo {author} {\bibfnamefont {H.~E.}\ \bibnamefont
  {Radford}}, \bibinfo {author} {\bibfnamefont {V.~W.}\ \bibnamefont {Hughes}},
  \ and\ \bibinfo {author} {\bibfnamefont {V.}~\bibnamefont {Beltran-Lopez}},\
  }\Doi {10.1103/PhysRev.123.153} {\bibfield  {journal} {\bibinfo  {journal}
  {Phys. Rev.},\ }\textbf {\bibinfo {volume} {123}},\ \bibinfo {pages} {153}
  (\bibinfo {year} {1961})}\BibitemShut {NoStop}%
\bibitem [{\citenamefont {Harvey}(1965)}]{Har:65a}%
  \BibitemOpen
  \bibfield  {author} {\bibinfo {author} {\bibfnamefont {J.~S.~M.}\
  \bibnamefont {Harvey}},\ }\Doi {10.1098/rspa.1965.0126} {\bibfield  {journal}
  {\bibinfo  {journal} {Proceedings of the Royal Society of London. Series A.
  Mathematical and Physical Sciences},\ }\textbf {\bibinfo {volume} {285}},\
  \bibinfo {pages} {581} (\bibinfo {year} {1965})},\ \Eprint
  {http://arxiv.org/abs/http://rspa.royalsocietypublishing.org/content/285/1403/581.full.pdf+html}
  {http://rspa.royalsocietypublishing.org/content/285/1403/581.full.pdf+html}
  \BibitemShut {NoStop}%
\bibitem [{\citenamefont {Glass}\ and\ \citenamefont
  {Hibbert}(1978)}]{GlaHib:78b}%
  \BibitemOpen
  \bibfield  {author} {\bibinfo {author} {\bibfnamefont {R.}~\bibnamefont
  {Glass}}\ and\ \bibinfo {author} {\bibfnamefont {A.}~\bibnamefont
  {Hibbert}},\ }\href {http://stacks.iop.org/0022-3700/11/i=13/a=009}
  {\bibfield  {journal} {\bibinfo  {journal} {J. Phys. B: Atom. Molec.
  Phys.,},\ }\textbf {\bibinfo {volume} {11}},\ \bibinfo {pages} {2257}
  (\bibinfo {year} {1978})}\BibitemShut {NoStop}%
\bibitem [{\citenamefont {Lid\'en}(1949)}]{Lid:49a}%
  \BibitemOpen
  \bibfield  {author} {\bibinfo {author} {\bibfnamefont {K.}~\bibnamefont
  {Lid\'en}},\ }\href@noop {} {\bibfield  {journal} {\bibinfo  {journal} {Ark.
  Fys.},\ }\textbf {\bibinfo {volume} {1}},\ \bibinfo {pages} {229} (\bibinfo
  {year} {1949})}\BibitemShut {NoStop}%
\bibitem [{\citenamefont {Hocker}(1978)}]{Hoc:78a}%
  \BibitemOpen
  \bibfield  {author} {\bibinfo {author} {\bibfnamefont {L.}~\bibnamefont
  {Hocker}},\ }\href@noop {} {\bibfield  {journal} {\bibinfo  {journal} {J.
  Opt. Soc. America},\ }\textbf {\bibinfo {volume} {68}},\ \bibinfo {pages}
  {262} (\bibinfo {year} {1978})}\BibitemShut {NoStop}%
\bibitem [{\citenamefont {Levy}\ \emph {et~al.}(2007)\citenamefont {Levy},
  \citenamefont {Cocolios}, \citenamefont {Behr}, \citenamefont {Jayamanna},
  \citenamefont {Minamisono},\ and\ \citenamefont {Pearson}}]{Levetal:07a}%
  \BibitemOpen
  \bibfield  {author} {\bibinfo {author} {\bibfnamefont {C.}~\bibnamefont
  {Levy}}, \bibinfo {author} {\bibfnamefont {T.}~\bibnamefont {Cocolios}},
  \bibinfo {author} {\bibfnamefont {J.}~\bibnamefont {Behr}}, \bibinfo {author}
  {\bibfnamefont {K.}~\bibnamefont {Jayamanna}}, \bibinfo {author}
  {\bibfnamefont {K.}~\bibnamefont {Minamisono}}, \ and\ \bibinfo {author}
  {\bibfnamefont {M.}~\bibnamefont {Pearson}},\ }\Doi
  {http://dx.doi.org/10.1016/j.nima.2007.07.013} {\bibfield  {journal}
  {\bibinfo  {journal} {Nuclear Instruments and Methods in Physics Research
  Section A: Accelerators, Spectrometers, Detectors and Associated Equipment},\
  }\textbf {\bibinfo {volume} {580}},\ \bibinfo {pages} {1571 } (\bibinfo
  {year} {2007})},\ ISSN \bibinfo {issn} {0168-9002}\BibitemShut {NoStop}%
\bibitem [{\citenamefont {Huo}\ \emph {et~al.}(2018)\citenamefont {Huo},
  \citenamefont {Deng}, \citenamefont {Windholz}, \citenamefont {Mu},\ and\
  \citenamefont {Wang}}]{Huoetal:2018a}%
  \BibitemOpen
  \bibfield  {author} {\bibinfo {author} {\bibfnamefont {X.}~\bibnamefont
  {Huo}}, \bibinfo {author} {\bibfnamefont {L.}~\bibnamefont {Deng}}, \bibinfo
  {author} {\bibfnamefont {L.}~\bibnamefont {Windholz}}, \bibinfo {author}
  {\bibfnamefont {X.}~\bibnamefont {Mu}}, \ and\ \bibinfo {author}
  {\bibfnamefont {H.}~\bibnamefont {Wang}},\ }\Doi
  {https://doi.org/10.1016/j.jqsrt.2017.09.020} {\bibfield  {journal} {\bibinfo
   {journal} {Journal of Quantitative Spectroscopy and Radiative Transfer},\
  }\textbf {\bibinfo {volume} {205}},\ \bibinfo {pages} {1 } (\bibinfo {year}
  {2018})},\ ISSN \bibinfo {issn} {0022-4073}\BibitemShut {NoStop}%
\bibitem [{\citenamefont {Marin}\ \emph {et~al.}(1993)\citenamefont {Marin},
  \citenamefont {Fort}, \citenamefont {Prevedelli}, \citenamefont {Inguscio},
  \citenamefont {Tino},\ and\ \citenamefont {Bauche}}]{Maretal:93a}%
  \BibitemOpen
  \bibfield  {author} {\bibinfo {author} {\bibfnamefont {F.}~\bibnamefont
  {Marin}}, \bibinfo {author} {\bibfnamefont {C.}~\bibnamefont {Fort}},
  \bibinfo {author} {\bibfnamefont {M.}~\bibnamefont {Prevedelli}}, \bibinfo
  {author} {\bibfnamefont {M.}~\bibnamefont {Inguscio}}, \bibinfo {author}
  {\bibfnamefont {G.}~\bibnamefont {Tino}}, \ and\ \bibinfo {author}
  {\bibfnamefont {J.}~\bibnamefont {Bauche}},\ }\href@noop {} {\bibfield
  {journal} {\bibinfo  {journal} {Z. Phys.D--Atoms, Molecules and Clusters},\
  }\textbf {\bibinfo {volume} {25}},\ \bibinfo {pages} {191} (\bibinfo {year}
  {1993})}\BibitemShut {NoStop}%
\bibitem [{\citenamefont {Godefroid}\ \emph {et~al.}(1997)\citenamefont
  {Godefroid}, \citenamefont {{Van Meulebeke}}, \citenamefont {J\"onsson},\
  and\ \citenamefont {{Froese Fischer}}}]{Godetal:97b}%
  \BibitemOpen
  \bibfield  {author} {\bibinfo {author} {\bibfnamefont {M.}~\bibnamefont
  {Godefroid}}, \bibinfo {author} {\bibfnamefont {G.}~\bibnamefont {{Van
  Meulebeke}}}, \bibinfo {author} {\bibfnamefont {P.}~\bibnamefont
  {J\"onsson}}, \ and\ \bibinfo {author} {\bibfnamefont {C.}~\bibnamefont
  {{Froese Fischer}}},\ }\href@noop {} {\bibfield  {journal} {\bibinfo
  {journal} {Z. Phys.D--Atoms, Molecules and Clusters},\ }\textbf {\bibinfo
  {volume} {42}},\ \bibinfo {pages} {193} (\bibinfo {year} {1997})}\BibitemShut
  {NoStop}%
\bibitem [{\citenamefont {J\"onsson}\ and\ \citenamefont
  {Godefroid}(2000)}]{JonGod:00a}%
  \BibitemOpen
  \bibfield  {author} {\bibinfo {author} {\bibfnamefont {P.}~\bibnamefont
  {J\"onsson}}\ and\ \bibinfo {author} {\bibfnamefont {M.}~\bibnamefont
  {Godefroid}},\ }\href@noop {} {\bibfield  {journal} {\bibinfo  {journal}
  {Molecular Physics},\ }\textbf {\bibinfo {volume} {98}},\ \bibinfo {pages}
  {1141} (\bibinfo {year} {2000})}\BibitemShut {NoStop}%
\bibitem [{\citenamefont {{Froese Fischer}}\ \emph {et~al.}(2016)\citenamefont
  {{Froese Fischer}}, \citenamefont {Godefroid}, \citenamefont {Brage},
  \citenamefont {J\"onsson},\ and\ \citenamefont {Gaigalas}}]{Froetal:2016a}%
  \BibitemOpen
  \bibfield  {author} {\bibinfo {author} {\bibfnamefont {C.}~\bibnamefont
  {{Froese Fischer}}}, \bibinfo {author} {\bibfnamefont {M.}~\bibnamefont
  {Godefroid}}, \bibinfo {author} {\bibfnamefont {T.}~\bibnamefont {Brage}},
  \bibinfo {author} {\bibfnamefont {P.}~\bibnamefont {J\"onsson}}, \ and\
  \bibinfo {author} {\bibfnamefont {G.}~\bibnamefont {Gaigalas}},\ }\href
  {http://stacks.iop.org/0953-4075/49/i=18/a=182004} {\bibfield  {journal}
  {\bibinfo  {journal} {J. Phys. B: At. Mol. Opt. Phys.},\ }\textbf {\bibinfo
  {volume} {49}},\ \bibinfo {pages} {182004} (\bibinfo {year}
  {2016})}\BibitemShut {NoStop}%
\bibitem [{\citenamefont {Lindgren}\ and\ \citenamefont
  {Ros\'en}(1975)}]{LINROS:75a}%
  \BibitemOpen
  \bibfield  {author} {\bibinfo {author} {\bibfnamefont {I.}~\bibnamefont
  {Lindgren}}\ and\ \bibinfo {author} {\bibfnamefont {A.}~\bibnamefont
  {Ros\'en}},\ }in\ \Doi {https://doi.org/10.1016/B978-0-7204-0331-2.50007-X}
  {\emph {\bibinfo {booktitle} {Case Studies in Atomic Physics}}},\ \bibinfo
  {editor} {edited by\ \bibinfo {editor} {\bibfnamefont {E.}~\bibnamefont
  {McDANIEL}}\ and\ \bibinfo {editor} {\bibfnamefont {M.}~\bibnamefont
  {McDOWELL}}}\ (\bibinfo  {publisher} {Elsevier},\ \bibinfo {year} {1975})\
  pp.\ \bibinfo {pages} {93 -- 196},\ ISBN \bibinfo {isbn}
  {978-0-7204-0331-2}\BibitemShut {NoStop}%
\bibitem [{\citenamefont {Hibbert}(1975)}]{Hib:75b}%
  \BibitemOpen
  \bibfield  {author} {\bibinfo {author} {\bibfnamefont {A.}~\bibnamefont
  {Hibbert}},\ }\href@noop {} {\bibfield  {journal} {\bibinfo  {journal} {Rep.
  Prog. Phys.},\ }\textbf {\bibinfo {volume} {38}},\ \bibinfo {pages} {1217}
  (\bibinfo {year} {1975})}\BibitemShut {NoStop}%
\bibitem [{\citenamefont {J\"onsson}\ \emph {et~al.}(1993)\citenamefont
  {J\"onsson}, \citenamefont {Wahlstr\"om},\ and\ \citenamefont
  {Fischer}}]{Jonetal:93a}%
  \BibitemOpen
  \bibfield  {author} {\bibinfo {author} {\bibfnamefont {P.}~\bibnamefont
  {J\"onsson}}, \bibinfo {author} {\bibfnamefont {C.-G.}\ \bibnamefont
  {Wahlstr\"om}}, \ and\ \bibinfo {author} {\bibfnamefont {C.~F.}\ \bibnamefont
  {Fischer}},\ }\href@noop {} {\bibfield  {journal} {\bibinfo  {journal} {Comp.
  Phys. Comm.},\ }\textbf {\bibinfo {volume} {74}},\ \bibinfo {pages} {399}
  (\bibinfo {year} {1993})}\BibitemShut {NoStop}%
\bibitem [{\citenamefont {{Froese Fischer}}\ \emph {et~al.}(1997)\citenamefont
  {{Froese Fischer}}, \citenamefont {Brage},\ and\ \citenamefont
  {J\"{o}nsson}}]{FBJ:97a}%
  \BibitemOpen
  \bibfield  {author} {\bibinfo {author} {\bibfnamefont {C.}~\bibnamefont
  {{Froese Fischer}}}, \bibinfo {author} {\bibfnamefont {T.}~\bibnamefont
  {Brage}}, \ and\ \bibinfo {author} {\bibfnamefont {P.}~\bibnamefont
  {J\"{o}nsson}},\ }\href@noop {} {\emph {\bibinfo {title} {Computational
  Atomic Structure - An MCHF Approach}}}\ (\bibinfo  {publisher} {Institute of
  Physics Publishing},\ \bibinfo {address} {Bristol},\ \bibinfo {year}
  {1997})\BibitemShut {NoStop}%
\bibitem [{\citenamefont {Froese~Fischer}\ \emph {et~al.}(2007)\citenamefont
  {Froese~Fischer}, \citenamefont {Tachiev}, \citenamefont {Gaigalas},\ and\
  \citenamefont {Godefroid}}]{Froetal:2007a}%
  \BibitemOpen
  \bibfield  {author} {\bibinfo {author} {\bibfnamefont {C.}~\bibnamefont
  {Froese~Fischer}}, \bibinfo {author} {\bibfnamefont {G.}~\bibnamefont
  {Tachiev}}, \bibinfo {author} {\bibfnamefont {G.}~\bibnamefont {Gaigalas}}, \
  and\ \bibinfo {author} {\bibfnamefont {M.}~\bibnamefont {Godefroid}},\
  }\href@noop {} {\bibfield  {journal} {\bibinfo  {journal} {Comput. Phys.
  Comm.},\ }\textbf {\bibinfo {volume} {176}},\ \bibinfo {pages} {559}
  (\bibinfo {year} {2007})}\BibitemShut {NoStop}%
\bibitem [{\citenamefont {{Armstrong Jr.}}\ and\ \citenamefont
  {Feneuille}(1974)}]{ArmFen:74a}%
  \BibitemOpen
  \bibfield  {author} {\bibinfo {author} {\bibfnamefont {L.}~\bibnamefont
  {{Armstrong Jr.}}}\ and\ \bibinfo {author} {\bibfnamefont {S.}~\bibnamefont
  {Feneuille}},\ }\href@noop {} {\bibfield  {journal} {\bibinfo  {journal}
  {Adv. At. Mol. Phys.},\ }\textbf {\bibinfo {volume} {10}},\ \bibinfo {pages}
  {1} (\bibinfo {year} {1974})}\BibitemShut {NoStop}%
\bibitem [{\citenamefont {J\"onsson}\ \emph {et~al.}(2013)\citenamefont
  {J\"onsson}, \citenamefont {Gaigalas}, \citenamefont {{Biero\'{n}}},
  \citenamefont {{Froese Fischer}},\ and\ \citenamefont
  {Grant}}]{Jonetal:2013a}%
  \BibitemOpen
  \bibfield  {author} {\bibinfo {author} {\bibfnamefont {P.}~\bibnamefont
  {J\"onsson}}, \bibinfo {author} {\bibfnamefont {G.}~\bibnamefont {Gaigalas}},
  \bibinfo {author} {\bibfnamefont {J.}~\bibnamefont {{Biero\'{n}}}}, \bibinfo
  {author} {\bibfnamefont {C.}~\bibnamefont {{Froese Fischer}}}, \ and\
  \bibinfo {author} {\bibfnamefont {I.}~\bibnamefont {Grant}},\ }\href@noop {}
  {\bibfield  {journal} {\bibinfo  {journal} {Comput. Phys. Comm.},\ }\textbf
  {\bibinfo {volume} {184}},\ \bibinfo {pages} {2197} (\bibinfo {year}
  {2013})}\BibitemShut {NoStop}%
\bibitem [{\citenamefont {Stone}(2005)}]{Sto:2005a}%
  \BibitemOpen
  \bibfield  {author} {\bibinfo {author} {\bibfnamefont {N.}~\bibnamefont
  {Stone}},\ }\href@noop {} {\bibfield  {journal} {\bibinfo  {journal} {Atomic
  Data and Nuclear Data Tables},\ }\textbf {\bibinfo {volume} {90}},\ \bibinfo
  {pages} {75} (\bibinfo {year} {2005})}\BibitemShut {NoStop}%
\bibitem [{\citenamefont {Pyykk\"o}(2008)}]{Pyy:2008a}%
  \BibitemOpen
  \bibfield  {author} {\bibinfo {author} {\bibfnamefont {P.}~\bibnamefont
  {Pyykk\"o}},\ }\Doi {10.1080/00268970802018367} {\bibfield  {journal}
  {\bibinfo  {journal} {Molecular Physics},\ }\textbf {\bibinfo {volume}
  {106}},\ \bibinfo {pages} {1965} (\bibinfo {year} {2008})},\ \Eprint
  {http://arxiv.org/abs/http://dx.doi.org/10.1080/00268970802018367}
  {http://dx.doi.org/10.1080/00268970802018367} \BibitemShut {NoStop}%
\bibitem [{\citenamefont {Stone}(2016)}]{Sto:2016a}%
  \BibitemOpen
  \bibfield  {author} {\bibinfo {author} {\bibfnamefont {N.}~\bibnamefont
  {Stone}},\ }\Doi {https://doi.org/10.1016/j.adt.2015.12.002} {\bibfield
  {journal} {\bibinfo  {journal} {Atomic Data and Nuclear Data Tables},\
  }\textbf {\bibinfo {volume} {111-112}},\ \bibinfo {pages} {1 } (\bibinfo
  {year} {2016})},\ ISSN \bibinfo {issn} {0092-640X}\BibitemShut {NoStop}%
\bibitem [{\citenamefont {Ruiz}\ \emph {et~al.}()\citenamefont {Ruiz},
  \citenamefont {Binnersley}, \citenamefont {Billowes}, \citenamefont
  {Bissell}, \citenamefont {Blaum}, \citenamefont {Cocolios}, \citenamefont
  {{Day Goodacre}}, \citenamefont {{de Groote}}, \citenamefont
  {{Farooq-Smith}}, \citenamefont {Flanagan}, \citenamefont {Franchoo},
  \citenamefont {Geithner}, \citenamefont {Gins}, \citenamefont {Heylen},
  \citenamefont {Koszor\'us}, \citenamefont {Kowalska}, \citenamefont
  {Kr\"amer}, \citenamefont {Lynch}, \citenamefont {{Malbrunot-Ettenauer}},
  \citenamefont {Marsh}, \citenamefont {Neyens}, \citenamefont {Neugart},
  \citenamefont {N\"ortersh\"auser}, \citenamefont {Rothe}, \citenamefont
  {S\'anchez}, \citenamefont {Stroke}, \citenamefont {Vernon}, \citenamefont
  {Wendt}, \citenamefont {Wilkins}, \citenamefont {Xu}, \citenamefont {Yang},\
  and\ \citenamefont {Yordanov}}]{Ruietal:2016a}%
  \BibitemOpen
  \bibfield  {author} {\bibinfo {author} {\bibfnamefont {R.}~\bibnamefont
  {Ruiz}}, \bibinfo {author} {\bibfnamefont {C.}~\bibnamefont {Binnersley}},
  \bibinfo {author} {\bibfnamefont {J.}~\bibnamefont {Billowes}}, \bibinfo
  {author} {\bibfnamefont {M.}~\bibnamefont {Bissell}}, \bibinfo {author}
  {\bibfnamefont {K.}~\bibnamefont {Blaum}}, \bibinfo {author} {\bibfnamefont
  {T.}~\bibnamefont {Cocolios}}, \bibinfo {author} {\bibfnamefont
  {T.}~\bibnamefont {{Day Goodacre}}}, \bibinfo {author} {\bibfnamefont
  {R.}~\bibnamefont {{de Groote}}}, \bibinfo {author} {\bibfnamefont
  {G.}~\bibnamefont {{Farooq-Smith}}}, \bibinfo {author} {\bibfnamefont
  {K.}~\bibnamefont {Flanagan}}, \bibinfo {author} {\bibfnamefont
  {S.}~\bibnamefont {Franchoo}}, \bibinfo {author} {\bibfnamefont
  {W.}~\bibnamefont {Geithner}}, \bibinfo {author} {\bibfnamefont
  {W.}~\bibnamefont {Gins}}, \bibinfo {author} {\bibfnamefont {H.}~\bibnamefont
  {Heylen}}, \bibinfo {author} {\bibfnamefont {A.}~\bibnamefont {Koszor\'us}},
  \bibinfo {author} {\bibfnamefont {M.}~\bibnamefont {Kowalska}}, \bibinfo
  {author} {\bibfnamefont {J.}~\bibnamefont {Kr\"amer}}, \bibinfo {author}
  {\bibfnamefont {K.}~\bibnamefont {Lynch}}, \bibinfo {author} {\bibfnamefont
  {S.}~\bibnamefont {{Malbrunot-Ettenauer}}}, \bibinfo {author} {\bibfnamefont
  {B.}~\bibnamefont {Marsh}}, \bibinfo {author} {\bibfnamefont
  {G.}~\bibnamefont {Neyens}}, \bibinfo {author} {\bibfnamefont
  {R.}~\bibnamefont {Neugart}}, \bibinfo {author} {\bibfnamefont
  {W.}~\bibnamefont {N\"ortersh\"auser}}, \bibinfo {author} {\bibfnamefont
  {S.}~\bibnamefont {Rothe}}, \bibinfo {author} {\bibfnamefont
  {R.}~\bibnamefont {S\'anchez}}, \bibinfo {author} {\bibfnamefont
  {H.}~\bibnamefont {Stroke}}, \bibinfo {author} {\bibfnamefont
  {A.}~\bibnamefont {Vernon}}, \bibinfo {author} {\bibfnamefont
  {K.}~\bibnamefont {Wendt}}, \bibinfo {author} {\bibfnamefont
  {S.}~\bibnamefont {Wilkins}}, \bibinfo {author} {\bibfnamefont
  {Z.}~\bibnamefont {Xu}}, \bibinfo {author} {\bibfnamefont {X.}~\bibnamefont
  {Yang}}, \ and\ \bibinfo {author} {\bibfnamefont {D.}~\bibnamefont
  {Yordanov}},\ }\href@noop {} {\enquote {\bibinfo {title} {Toward laser
  spectroscopy of exotic fluorine isotopes},}\ }\bibinfo {note} {INTC-I-171 to
  the ISOLDE and Neutron Time-of-Flight Committee (European Organization for
  Nuclear Research, 2016); available at:
  https://cds.cern.ch/record/2157183/files/INTC-I-171.pdf}\BibitemShut
  {NoStop}%
\end{thebibliography}
%

\end{document}